\documentstyle[12pt,epsfig]{article}
                                                                                
\begin{document}                                                                
\date{}
                                                                                
\title{                                                                         
{\vspace{-1em} \normalsize                                                      
\hfill \parbox{50mm}{DESY 97-132}}\\[25mm]
Quadratically optimized polynomials for                      \\
fermion simulations                                          \\[12mm]}
\author{ I. Montvay                                          \\
Deutsches Elektronen-Synchrotron DESY,                       \\
Notkestr.\,85, D-22603 Hamburg, Germany}                                       
                                                                                
\newcommand{\be}{\begin{equation}}                                              
\newcommand{\ee}{\end{equation}}                                                
\newcommand{\half}{\frac{1}{2}}                                                 
\newcommand{\rar}{\rightarrow}                                                  
\newcommand{\lar}{\leftarrow}                                                   
                                                                                
\maketitle
\vspace{3em}
                                                                                
\begin{abstract} \normalsize
 Quadratically optimized polynomials are described which are useful in
 multi-bosonic algorithms for Monte Carlo simulations of quantum field
 theories with fermions.
 Algorithms for the computation of the coefficients and roots of
 these polynomials are described and their implementation in the
 algebraic manipulation language Maple is discussed.
 Tests of the evaluation of polynomials on dynamical fermion
 configurations are performed.
 In a simple special case the obtained polynomial approximations are
 compared to Chebyshev polynomials.
\end{abstract}       

\newpage
\section{Introduction}\label{sec1}
 The numerical simulation of quantum field theories with fermions is an
 interesting and difficult computational problem.
 The basic difficulty is the necessary calculation of very large
 determinants, the determinants of {\em fermion matrices}, which can
 only be achieved by some stochastic procedure with the help of
 auxiliary bosonic ``pseudofermion'' fields (for general background see,
 for instance, \cite{MONMUN}).

 A promising new approach proposed by L\"uscher \cite{LUSCHER} is based
 on polynomial approximations of some negative powers $x^{-\alpha}$ of
 the fermion matrix.
 In this case the number of auxiliary bosonic fields is equal to the
 order of the polynomial and the resulting {\em multi-bosonic action}
 can be treated by simple methods known from bosonic quantum field
 theories without fermions.
 The performance of such a {\em multi-bosonic algorithm} can be improved
 in a two-step polynomial approximation scheme \cite{TWO-STEP}, where
 the number of bosonic fields is equal to the order of a first,
 relatively low order, polynomial only realizing a modest approximation
 of $x^{-\alpha}$.
 The required high precision is achieved in a ``noisy'' correction step
 as proposed some time ago by Kennedy and Kuti \cite{NOISY}.
 In the correction step some high order polynomial approximations of
 $x^{-\alpha}/\bar{P}(x)$ are used, where $\bar{P}(x)$ is some
 polynomial.
 The necessary polynomials can be obtained by minimizing the integral
 of squared relative deviations in an interval containing the eigenvalue
 spectrum of the fermion matrix.

 The computation of the quadratically optimized polynomial
 approximations of functions of general type $x^{-\alpha}/\bar{P}(x)$,
 which are required in the two-step multi-bosonic algorithms, has been
 briefly outlined in ref.~\cite{TWO-STEP}.
 It can be best done by using the arbitrarily high precision facilities
 of an algebraic manipulation program, as for instance Maple V.
 In the present paper the algorithms for the calculation of the
 coefficients and roots of these polynomials are described in more
 detail.
 The optimal ordering of the roots in applying the polynomials of the
 fermion matrix in a product form is also discussed.
 This is necessary in order to keep rounding errors tolerable even for
 32 bit arithmetics.

 These kinds of quadratically optimized polynomial approximations
 belong to the general class of ``least-squares'' or ``Gaussian-''
 approximations \cite{CHEBYSHEV}.
 The coefficients of the polynomials can be obtained, for instance, by
 expanding the function to be approximated in terms of suitably defined
 orthogonal polynomials.
 The expansion in orthogonal polynomials provides the possibility of a
 recursive evaluation without knowing the roots, which is advantageous
 from the point of view of rounding errors.
 In case if the roots of the approximating polynomials are also
 required, the calculation of the polynomial coefficients is the
 smaller part of the calculation.
 The larger part is to determine the roots with sufficient precision
 and find their optimal ordering.

 Another possibility, besides the quadratic optimization, is to consider
 ``minimax'' or ``infinity-norm'' approximation schemes, where the
 maximum deviation in the region of approximation is minimized 
 \cite{CHEBYSHEV}.
 For instance, in ref.~\cite{LUSCHER} this has been used to approximate
 the function $x^{-1}$ by Chebyshev polynomials.
 In multi-bosonic algorithms, in general, the quadratic optimization
 seems more appealing, because it provides better approximations in
 the average.
 In addition, the advantage of the quadratic optimization is its great
 flexibility.
 For instance, besides the possibility to choose a wide class of
 functions, it also allows for arbitrary complex regions and
 general weight factors which also take into account the average
 eigenvalue distribution of the fermion matrix.
 Despite this generality, in all cases one can use essentially the same
 algorithm to determine the polynomials.
 Nevertheless, in the case of the function $x^{-1}$ the choice between
 minimax and least-squares approximation can only be decided by
 practical tests and the outcome may also depend on the preferences
 of a particular application.
 (The same applies to the generalization of the Chebyshev approximation
 to the more general case of $x^{-\alpha}$ which has been proposed
 recently by Bunk \cite{BUNK}.)

 The plan of this paper is as follows:
 In the next section first the definitions and some basic properties of
 the polynomials necessary in the first step of the two-step
 multi-bosonic algorithm for fermions are collected.
 In subsection~\ref{sec2.1} the questions of the expansion in orthogonal
 polynomials are discussed. 
 The Maple procedures for the calculation of quadratically optimized
 polynomials and for obtaining the optimal orderings of their roots are
 considered in section \ref{sec3}, together with a few examples.
 Possible generalizations of these results are discussed in section
 \ref{sec4}.
 These include the polynomials needed in the second (correction-) step
 of the fermion algorithm and also some other cases useful for other
 variants of multi-bosonic algorithms.
 Some tests are included in section~\ref{sec5}, together with
 comparisons to the minimax approximation in the special case of the
 function $x^{-1}$.
 The last section contains a summary and concluding remarks.
 

\section{Definitions and some basic properties}\label{sec2}
 Quadratically optimized approximations of a function $f(x)$ by 
 polynomials $P(x)$ in an interval $x \in [\epsilon,\lambda]$ can be
 defined by minimizing the integral
\be \label{eq01}                                                                
\int_\epsilon^\lambda dx \left[ (f(x) - P(x))w(x) \right]^2 \ .
\ee                                                                             
 Here $w(x)$ is an arbitrary weight function.
 Minimizing the relative deviation means to choose $w(x)=1/f(x)$.
 Here we shall first mainly consider some negative power
 $f(x) \equiv x^{-\alpha}$ and a positive interval
 $0 \leq \epsilon < \lambda$.
 (The required generalization to $f(x) \equiv x^{-\alpha}/\bar{P}(x)$
 and more general regions will be discussed in section~\ref{sec4}.)
 In the multi-bosonic algorithm for fermions we need $\alpha=\half N_f$,
 with $N_f$ the number of identical fermion flavours.
 For numerical simulations with gluinos, which are Majorana fermions,
 we have $N_f=\half$.
 In QCD $N_f=1,2,3$ are relevant, depending on the assumptions made on
 the values of quark masses.

 Therefore, we shall minimize the {\em relative deviation norm}
\be \label{eq02}
\delta \equiv \left\{ (\lambda - \epsilon)^{-1}
\int_\epsilon^\lambda dx \left[ 1 - x^\alpha P(x) \right]^2
\right\}^\half \ .
\ee
 The normalization factor in front of the integral is, of course, not
 necessary.
 It is introduced for convenience.
 Taking the square root is also a matter of convention.

 $\delta^2$ is a quadratic form in the coefficients of the polynomial
 which can be straightforwardly minimized.
 Let us denote the polynomial corresponding to the minimum of $\delta$
 by
\be \label{eq03}
P_n(\alpha;\epsilon,\lambda;x) \equiv
\sum_{\nu=0}^n c_{n\nu}(\alpha;\epsilon,\lambda) x^{n-\nu} \ .
\ee
 Performing the integral in $\delta^2$ term by term we obtain
\be \label{eq04}
\delta^2 = 1 -2 \sum_{\nu=0}^n c_\nu V_\nu^{(\alpha)}
+ \sum_{\nu_1,\nu_2=0}^n c_{\nu_1} M_{\nu_1,\nu_2}^{(\alpha)} c_{\nu_2}
\ ,
\ee
 where
$$
V_\nu^{(\alpha)} =
\frac{\lambda^{1+\alpha+n-\nu}-\epsilon^{1+\alpha+n-\nu}}
              {(\lambda - \epsilon)(1+\alpha+n-\nu)} \ ,
$$
\be \label{eq05} 
M_{\nu_1,\nu_2}^{(\alpha)} = 
\frac{\lambda^{1+2\alpha+2n-\nu_1-\nu_2} -
     \epsilon^{1+2\alpha+2n-\nu_1-\nu_2}}
              {(\lambda - \epsilon)(1+2\alpha+2n-\nu_1-\nu_2)}  \ .
\ee
 Note that the dependence of $V$ and $M$ on $n$ comes only from the
 dimensions.
 This can be made explicit by introducing in (\ref{eq03}) the
 coefficients $\tilde{c}_{n\nu} \equiv c_{n,n-\nu}$.
 Then everywhere in (\ref{eq05}) we replace $(n-\nu) \to \nu$.

 The coefficients of the polynomial corresponding to the minimum of
 $\delta^2$, or of $\delta$, are
\be \label{eq06}
c_\nu \equiv c_{n\nu}(\alpha;\epsilon,\lambda) =
\sum_{\nu_1=0}^n M_{\nu\nu_1}^{(\alpha)-1} V_{\nu_1}^{(\alpha)} \ . 
\ee
 The value at the minimum is
\be \label{eq07}
\delta^2 \equiv \delta_n^2(\alpha;\epsilon,\lambda) 
= 1 - \sum_{\nu_1,\nu_2=0}^n 
V_{\nu_1}^{(\alpha)} M_{\nu_1,\nu_2}^{(\alpha)-1} V_{\nu_2}^{(\alpha)}
\ .
\ee

 Scaling the integration variable $x$ by $x^\prime = \rho x$ one obtains
 the following scaling properties of the optimized polynomials:
$$
\delta_n^2(\alpha;\epsilon\rho,\lambda\rho)
= \delta_n^2(\alpha;\epsilon,\lambda) \ ,
$$
\be \label{eq08}
P_n(\alpha;\epsilon\rho,\lambda\rho;x) = \rho^{-\alpha}
P_n(\alpha;\epsilon,\lambda;x/\rho) \ ,
\hspace{3em}
c_{n\nu}(\alpha;\epsilon\rho,\lambda\rho)
= \rho^{n-\nu-\alpha} c_{n\nu}(\alpha;\epsilon,\lambda) \ .
\ee
 This allows for only considering, for instance, the standard intervals
 $[\epsilon/\lambda, 1]$ and obtain other cases by these scaling
 relations.

 In applications to multi-bosonic algorithms for fermions the
 decomposition of the optimized polynomials as a product of root-factors
 is needed.
 This can be written as
\be \label{eq09}
P_n(\alpha;\epsilon,\lambda;x) =
c_{n0}(\alpha;\epsilon,\lambda) \prod_{j=1}^n 
[x-r_{nj}(\alpha;\epsilon,\lambda)] \ . 
\ee
 The scaling properties here are:
\be \label{eq10}
c_{n0}(\alpha;\epsilon\rho,\lambda\rho)
=\rho^{n-\alpha} c_{n0}(\alpha;\epsilon,\lambda) 
\ , \hspace{3em}
r_{nj}(\alpha;\epsilon\rho,\lambda\rho)
= \rho^j r_{nj}(\alpha;\epsilon,\lambda) \ .
\ee
 The numerical calculation of roots will be discussed in the next
 section.
 Since the coefficients of the polynomials are real, the roots are
 either real or occur in complex conjugate pairs.
 In the present case, for $n=even$ the roots always occur in pairs with
 non-zero imaginary parts.
 For $n=odd$ there is a single real root above the upper limit of the
 interval $\lambda$, and the other roots are in complex conjugate pairs.
 
 The above formulae also apply in the limit $\epsilon \to 0$.
 Other generalizations will be discussed in section~\ref{sec4}.

 From the minimum property of the optimized polynomials and from the
 positivity of the integrand in eq.~(\ref{eq02}) one can derive
 interesting inequalities for the relative deviation norm $\delta$.
 For $0 \leq \epsilon^\prime < \epsilon$ we obtain
\be \label{eq11}
(1-\epsilon)\delta_n^2(\alpha;\epsilon,1)
\;<\; (1-\epsilon^\prime)\delta_n^2(\alpha;\epsilon^\prime,1) \ .
\ee
 Of course, for $m > n$  we also have 
 $\delta_m(\alpha;\epsilon,\lambda) < \delta_n(\alpha;\epsilon,\lambda)$.
 One can also prove
\be \label{eq12}
\delta_n^2(\alpha;0,1) -\epsilon
\;<\; (1-\epsilon)\delta_n^2(\alpha;\epsilon,1) \ ,
\ee
 and for integer $k \geq 2$ and large enough $n$
\be \label{eq13}
\delta_{kn}^2(k\alpha;\epsilon,\lambda)
\;<\; k^2 \delta_n^2(\alpha;\epsilon,\lambda) \ .
\ee

 In general, it does not seem to be possible to derive explicit formulae
 for $\delta$. In some special cases, however, explicit algebraic
 calculations in Maple give interesting illustrative results.
 For instance, in the important special case $\alpha=\half$, we have
$$
\delta_4^2(\half;\epsilon^2,1) =
$$
\be \label{eq14}
\frac{(\epsilon^{10}+20\epsilon^9+105\epsilon^8+320\epsilon^7
+580\epsilon^6+720\epsilon^5+580\epsilon^4+320\epsilon^3
+105\epsilon^2+20\epsilon+1)(\epsilon-1)^{10}}
{121 (\epsilon+1)^{10}
(\epsilon^8+24\epsilon^6+76\epsilon^4+24\epsilon^2+1)(\epsilon^2+1)} \ .
\ee
 In the limit $\epsilon \to 0$ we have, for any $\alpha$ and at least
 for a wide range of orders $n$ where explicit calculation could be
 performed,
\be \label{eq15}
\delta_n(\alpha;0,1) = \frac{\alpha}{n+1+\alpha} \ .
\ee
%


\subsection{Expansion in orthogonal polynomials}\label{sec2.1}
 A useful representation of the quadratically optimized polynomials is
 an expansion in suitably defined orthogonal polynomials.
 If the relative deviation from the function
 $f(x)=x^{-\alpha}/\bar{P}(x)$ is minimized, the appropriate integration
 weight in (\ref{eq01}) is
\be \label{eq16}
w(x)^2 = \frac{1}{f(x)^2} = \left[ x^\alpha \bar{P}(x) \right]^2 \ . 
\ee
 In most cases this is the preferred choice for the weight function.
 Nevertheless, also other cases are of interest, for instance $w(x)=1$
 and $f(x)=x^\alpha\bar{P}(x)$, as in eqs.~(58)-(59) of
 ref.~\cite{TWO-STEP}.
 Therefore, let us leave for the moment $w(x)$ and $f(x)$ general.

 Let us define the orthogonal polynomials $\Phi_\nu(x)$ of order
 $\nu=0,1,2,\ldots$ such that they satisfy
\be \label{eq17}
\int_\epsilon^\lambda dx\, w(x)^2 \Phi_\mu(x)\Phi_\nu(x) 
= \delta_{\mu\nu} q_\nu \ .
\ee
 The arbitrary normalization factor $q_\nu$ will be chosen later on by
 convenience.
 The orthogonal polynomials can be used, instead of the simple powers in
 (\ref{eq03}), as a basis for the expansion of the optimized
 polynomials:
\be \label{eq18}
P_n(\alpha;\epsilon,\lambda;x) =
\sum_{\nu=0}^n d_{n\nu}(\alpha;\epsilon,\lambda) \Phi_\nu(x) \ .
\ee
 The advantage of this expansion is that now, as one can easily see,
 the matrix corresponding to $M^{(\alpha)}$ in (\ref{eq05}) is diagonal
 and the coefficients of the optimized polynomial are simply given by
\be \label{eq19}
d_{n\nu}(\alpha;\epsilon,\lambda) \equiv d_n(\alpha;\epsilon,\lambda)
= \frac{b_\nu}{q_\nu} \ ,
\ee
 where
\be \label{eq20}
b_\nu \equiv \int_\epsilon^\lambda dx\, w(x)^2 f(x) \Phi_\nu(x) \ .
\ee
 An important property of this expansion is that, as shown by the
 notation, the expansion coefficients $d_{n\nu}=d_\nu$ do not depend on
 the order of the optimized polynomial $n$.

 The relations in eqs.~(\ref{eq17})-(\ref{eq20}) show that (\ref{eq18})
 is the truncated expansion of the function $f(x)$ in terms of the basis
 defined by $\Phi_\nu(x)$ in the Hilbert space of square integrable
 functions in the interval $[\epsilon,\lambda]$ with integration measure
 $dx\; w(x)^2$.
 A general consequence is that for $n \to \infty$ the approximation
 is exponential, if the function is bounded as, for instance,
 $f(x) = x^{-\alpha}/\bar{P}(x)$ for $\epsilon > 0$.

 One can easily show that the minimum of the relative deviation norm
 in (\ref{eq07}), for simplicity in the case $w(x)=1/f(x)$, can now be
 expressed as
$$
\delta_n^2 = (\lambda-\epsilon)^{-1} \int_\epsilon^\lambda dx\, 
\left[ 1 - w(x)\sum_{\nu=0}^n d_\nu\Phi_\nu(x) \right]^2
$$
\be \label{eq21}
= (\lambda-\epsilon)^{-1} \int_\epsilon^\lambda dx\,
\left[ 1 - w(x)\sum_{\nu=0}^n d_\nu\Phi_\nu(x) \right]
= 1 - (\lambda-\epsilon)^{-1}\sum_{\nu=0}^n d_\nu b_\nu \ .
\ee

 For the determination of the orthogonal polynomials $\Phi_\nu(x)$
 one can use recurrence relations \cite{CHEBYSHEV}.
 Let us define the expansion of $\Phi_\mu(x)$ in simple powers by
\be \label{eq22}
\Phi_\mu(x) = \sum_{\nu=0}^\mu f_{\mu\nu}x^{\mu-\nu} \ .
\ee
 The up to now arbitrary normalization factor $q_\nu$ in (\ref{eq17})
 can be fixed by requiring
\be \label{eq23}
f_{\mu 0} = 1 \ , \hspace{3em} (\mu=0,1,2,\ldots) \ .
\ee
 The first two polynomials with $\mu=0,1$ are:
\be \label{eq24}
\Phi_0(x) = 1 \ , \hspace{3em}
\Phi_1(x) = x + f_{11} = x - \frac{s_1}{s_0} \ ,
\ee
 with the notation
\be \label{eq25}
s_\mu \equiv \int_\epsilon^\lambda dx\, w(x)^2 x^\mu \ .
\ee
 The value of the coefficient $f_{11}$ is determined by the requirement
 of orthogonality of $\Phi_0(x)$ and $\Phi_1(x)$.
 The normalization factors $q_0$ and $q_1$ are given by
\be \label{eq26}
q_0 = s_0 \ , \hspace{3em} q_1 = s_2 - \frac{s_1^2}{s_0} \ .
\ee

 The higher order polynomials $\Phi_\mu(x)$ for $\mu=2,3,\ldots$ can be
 obtained from the three-term recurrence relation
\be \label{eq27}
\Phi_{r+1}(x) = (x+\beta_r)\Phi_r(x) + \gamma_{r-1}\Phi_{r-1}(x) \ .
\hspace{3em} (r=1,2,\ldots) \ .
\ee
 This relation follows from the fact that $\Phi_{r+1}(x)-x\Phi_r(x)$
 is a polynomial of order $r$, which is orthogonal to
 $\Phi_{r-2}(x),\;\Phi_{r-3}(x),\;\ldots$.
 Multiplying eq.~(\ref{eq27}) by $w(x)^2\Phi_r(x)$ and
 $w(x)^2\Phi_{r-1}(x)$ and integrating for $x \in [\epsilon,\lambda]$ 
 we obtain for the recurrence coefficients
\be \label{eq28}
\beta_r = -\frac{p_r}{q_r} \ , \hspace{3em}
\gamma_{r-1} = -\frac{q_r}{q_{r-1}} \ ,
\ee
 where
\be \label{eq29}
p_\nu \equiv \int_\epsilon^\lambda dx\, w(x)^2 \Phi_\nu(x)^2 x \ .
\ee
 Combining eqs.~(\ref{eq17}) and (\ref{eq29}) with (\ref{eq22}) and
 (\ref{eq25}) we also obtain
\be \label{eq30}
p_\mu = \sum_{\nu_1,\nu_2=0}^\mu f_{\mu\nu_1}f_{\mu\nu_2} 
s_{1+2\mu-\nu_1-\nu_2} \ , \hspace{3em}
q_\mu = \sum_{\nu_1,\nu_2=0}^\mu f_{\mu\nu_1}f_{\mu\nu_2} 
s_{  2\mu-\nu_1-\nu_2} \ .
\ee

 From the recurrence relation (\ref{eq27}), together with
 eqs.~(\ref{eq28}) and (\ref{eq30}), one can recursively determine the
 coefficients of the orthogonal polynomials $f_{\mu\nu}$.
 Calculating the expansion coefficients of the quadratically optimized
 polynomials from eqs.~(\ref{eq18})-(\ref{eq20}) one obtains the
 desired polynomial approximation, without the knowledge of roots.
 Besides the representation by a product of root-factors in
 (\ref{eq09}), this offers an alternative recursive way for the
 evaluation of optimized polynomials.


\section{Algorithm in Maple and examples}\label{sec3}
 The calculation of the coefficients of optimized polynomials and their
 roots from the formulae of the previous section is, in principle,
 straightforward.
 The problem is, however, that for high orders $n={\cal O}(100)$ the
 rounding errors in floating point calculations become dangerous.
 This holds for the determination of the inverse matrix $M^{(\alpha)-1}$
 in eq.~(\ref{eq06}), for finding the roots of the polynomial required
 in (\ref{eq09}) and for obtaining the coefficients in the recursion
 relation in eq.~(\ref{eq27}).
 Once the roots are found to sufficient precision, another problem is
 that applying the product representation (\ref{eq09}), with $x$
 replaced by the (squared) fermion matrix, precision losses occur again.
 This is because the intermediate results can be widely different in
 magnitude for different parts of the eigenvalue spectrum.
 A remedy is to find an optimized ordering of roots for the product
 of root-factors.
 All these problems can be dealt with by procedures written in an
 algebraic manipulation program as Maple V.
\begin{itemize}
\item{\bf Inversion and finding the roots:}
 The inversion of the matrix defined in (\ref{eq05}) can be done by the
 procedure {\em linalg[inverse]} included in the Maple V library.
 For the calculation of roots it is more convenient to write a special
 procedure rather than to use the Maple V library.
 The Laguerre iteration algorithm \cite{LAGUERRE} is well suited and
 can be easily implemented.
 In the examples explicitly considered in this paper the coefficients
 of the polynomials are real.
 This could, in principle, be used to accelerate the root-finding
 algorithm but, for keeping the procedure general, it is better not to
 use the realness of the coefficients.
 This also allows for checking the precision by looking whether the
 roots are either real or occur in complex conjugate pairs, as they
 should.
 The necessary number of digits for calculating the roots is usually
 smaller than $2n$.
 It does not depend much on the value of the power $\alpha$.
 (In the tests mainly $\alpha=1,\half,{1 \over 4}$ have been
 considered.)
 The precision requirements for matrix inversion and for finding the
 roots are roughly similar.

 The calculation on typical workstations for the interesting orders
 up to, say, $n=100$ can be performed within several hours with
 storage of several 10 Mbytes.
 Also higher orders with $n$ equal to several hundreds are feasible
 with reasonable computational effort: one has to have in mind that
 once the optimized polynomials are found they will typically be used
 in fermionic Monte Carlo simulation runs lasting up to several months
 on the largest supercomputers.
\item{\bf Optimized ordering of roots:}
 The product representation (\ref{eq09}), with $x$ replaced by the
 squared fermion matrix $X^2$, has to be applied to some vectors
 during the Monte Carlo simulations.
 This has to be done with care because of possible precision losses.
 In order to allow for a correct evaluation with 64-bit (or even
 32-bit) floating point arithmetics, one has to choose the ordering of
 the factors judiciously.
 Since in the intermediate steps of an iterative evaluation the vector
 components are collected as linear combinations of several terms,
 precision losses can occur when different parts of the eigenvalue
 spectrum are multiplied by widely different values of the partial
 products.
 This can be to a large extent avoided by optimizing the ordering of
 roots.

 It turned out that a good choice is to minimize the maximal ratio of
 the values $x^\alpha P_p(x)$ for $x \in [\epsilon,\lambda]$, with
 $P_p(x)$ denoting the value of the partial product under consideration.
 In practice this can be achieved by chosing a discrete number of
 points $\{x_1,x_2,\ldots,x_N\}$ in the interval
 $[\epsilon,\lambda]$ and, in a given step, comparing the values of
 $x^\alpha P_o(x)(x-r_j)$ for different choices of the next $r_j$.
 ($P_o(x)$ is the optimized partial product obtained in the previous
 step.)
 The next root $r_j$ is chosen such that the maximal ratio of the
 values of $x^\alpha P_o(x)(x-r_j)$ for $x \in \{x_1,x_2,\ldots,x_N\}$
 be minimal.
 The number of equidistant optimization points has been chosen
 typically to be $N={\cal O}(n)$.
 For $\alpha=\half,{1 \over 4}$ and $n \simeq 100-200$ even
 $N\simeq n/3 - n/5$ turned out to be sufficient.

 For some purposes it is better to hold complex conjugate root pairs
 together in the ordering.
 This is the case, for instance, if the splitting up of the polynomial
 in two complex conjugate halves is used, as discussed in
 section~\ref{sec5}.
 This ordering is somewhat worse for the whole polynomial, but it is
 necessary if the evaluation of the half-polynomials is required.

 An important characteristics of this ordering principle is that the
 choice of the next root depends on the previously chosen ones.
 Generalizations are possible, for instance, to decide on the basis
 of the maximal ratio in the next two or three or more steps, not only
 in the next step alone.
 In this way the choice of the next root will also depend on the later
 choices and the optimization develops a global character.
\item{\bf Recursion coefficients:}
 The recurrence relation (\ref{eq27}) requires to calculation of the
 coefficients $\beta_r,\gamma_{r-1}$.
 For this one first determines the values of the basic integrals
 $s_\mu$ in (\ref{eq25}) and then uses eqs.~(\ref{eq28})-(\ref{eq30}).
 In this case the necessary number of digits is also high.
 For the simple functions $x^{-\alpha}$ it is somewhat lower than
 for the determination of roots, but in the more general case
 $x^{-\alpha}/\bar{P}(x)$ the requirements are practically the same.
\end{itemize} 

\begin{figure}
\begin{center}
\epsfig{file=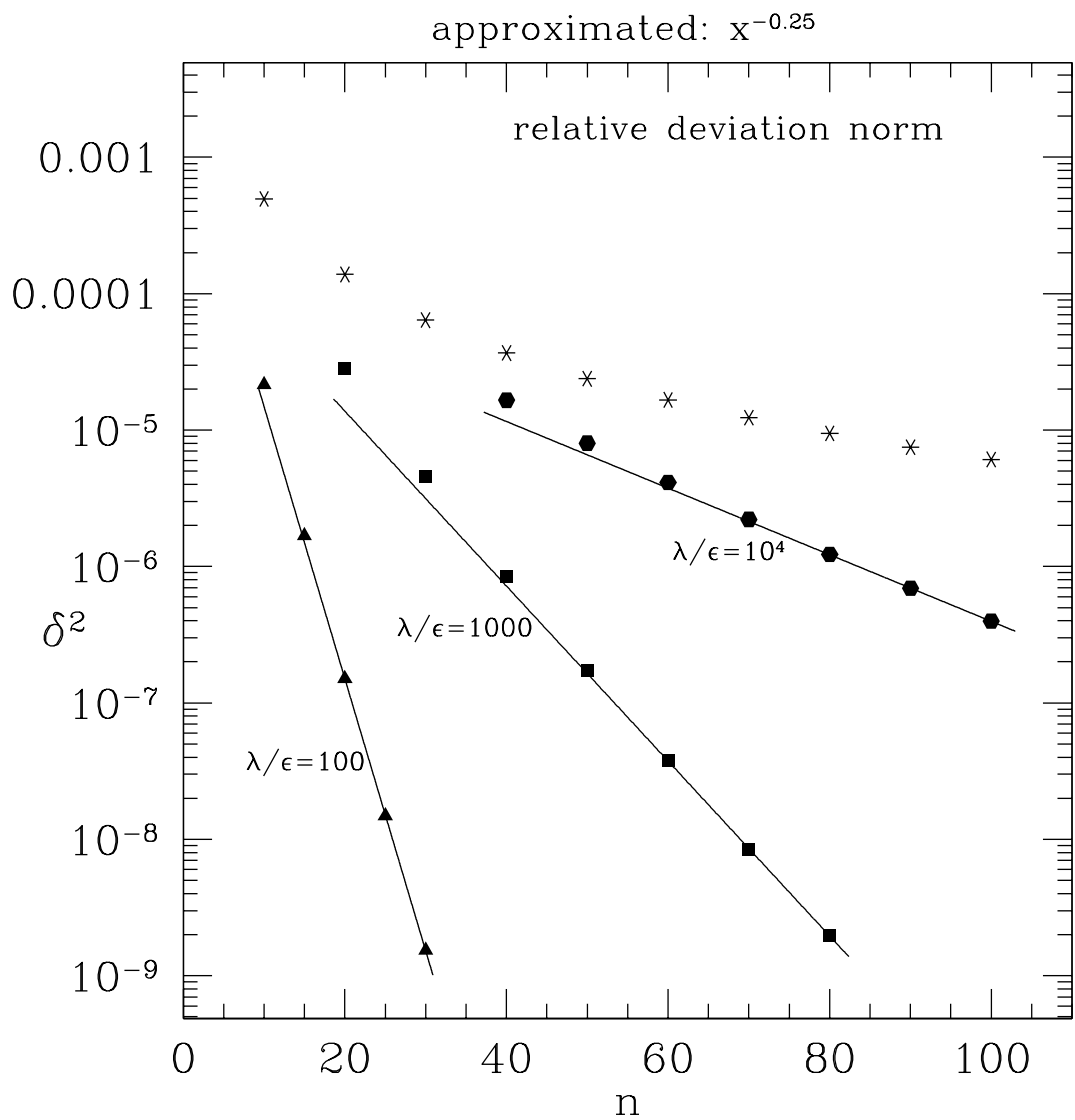,
        width=15.0cm,height=15.0cm,
        bbllx=20pt,bblly=150pt,bburx=600pt,bbury=720pt,
        angle=0}
\end{center}
\vspace{-1.0em}
\begin{center}
\parbox{15cm}{\caption{ \label{fig01}
 The (squared) deviation norm $\delta^2$ of the polynomial
 approximations of $x^{-1/4}$ as function of the order for different
 values of $\lambda/\epsilon$.
 The straight lines are fits to the last three points.
 The asterisks show the $\epsilon/\lambda \to 0$ limit given by
 eq.~({\protect\ref{eq15}}).}}
\end{center}
\end{figure}
\begin{figure}
\begin{center}
\epsfig{file=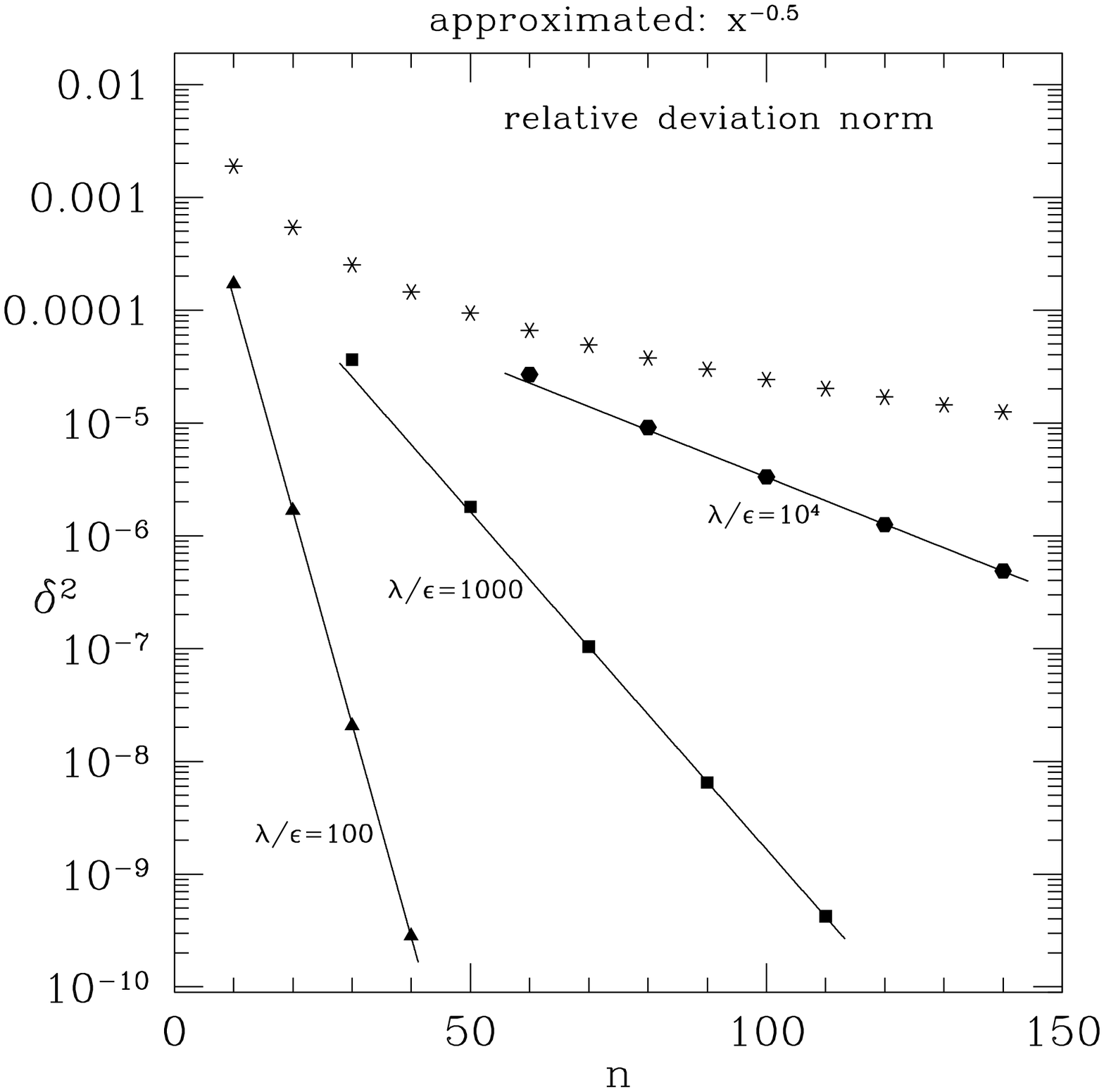,
        width=15.0cm,height=15.0cm,
        bbllx=20pt,bblly=150pt,bburx=600pt,bbury=720pt,
        angle=0}
\end{center}
\vspace{-1.0em}
\begin{center}
\parbox{15cm}{\caption{ \label{fig02}
 The (squared) deviation norm $\delta^2$ of the polynomial
 approximations of $x^{-1/2}$ as function of the order for different
 values of $\lambda/\epsilon$.
 The straight lines are fits to the last three points.
 The asterisks show the $\epsilon/\lambda \to 0$ limit given by
 eq.~({\protect\ref{eq15}}).}}
\end{center}
\end{figure}
\begin{figure}
\begin{center}
\epsfig{file=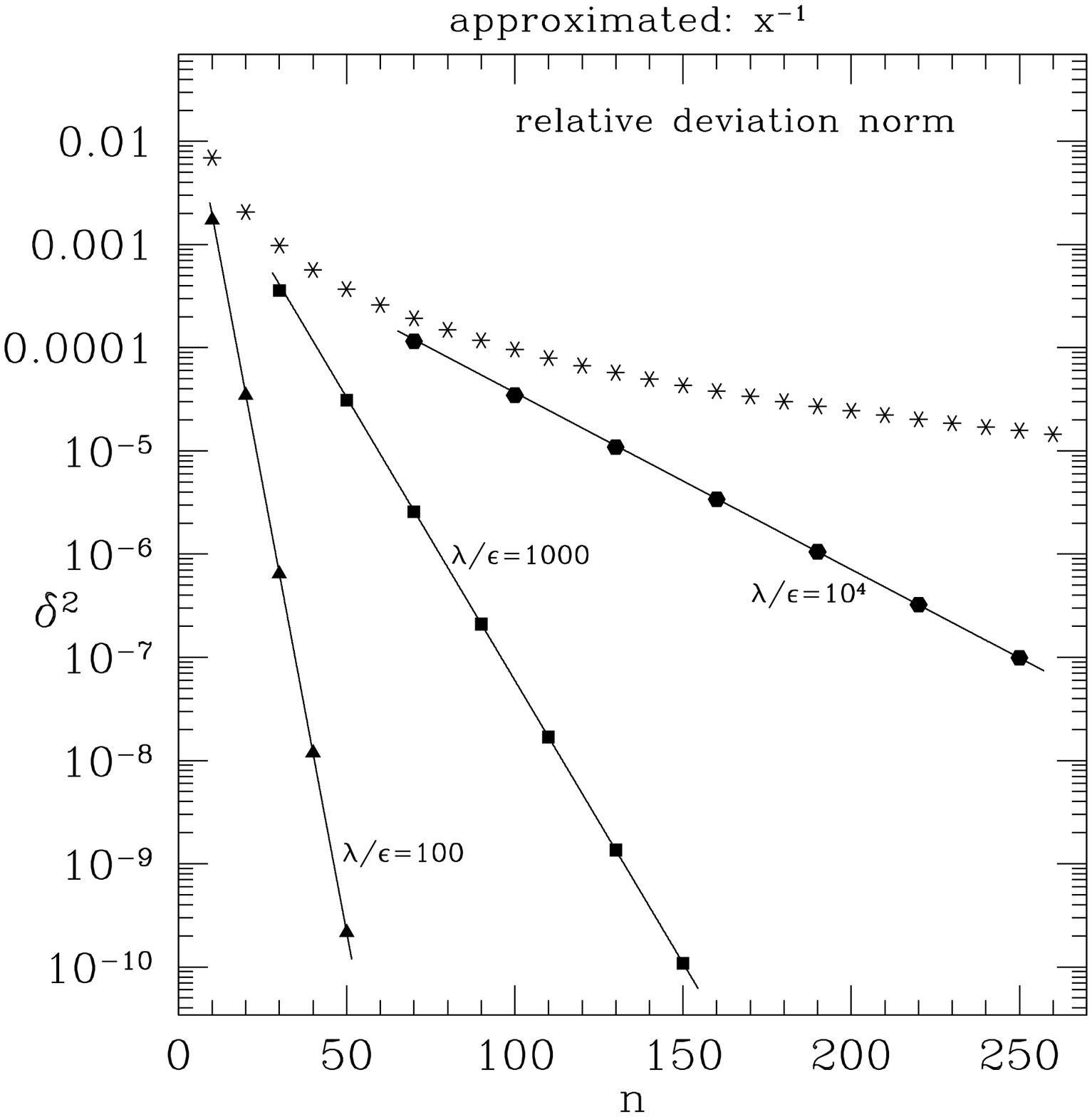,
        width=15.0cm,height=15.0cm,
        bbllx=20pt,bblly=150pt,bburx=600pt,bbury=720pt,
        angle=0}
\end{center}
\vspace{-1.0em}
\begin{center}
\parbox{15cm}{\caption{ \label{fig03}
 The (squared) deviation norm $\delta^2$ of the polynomial
 approximations of $x^{-1}$ as function of the order for different
 values of $\lambda/\epsilon$.
 The straight lines are fits to the last three points.
 The asterisks show the $\epsilon/\lambda \to 0$ limit given by
 eq.~({\protect\ref{eq15}}).}}
\end{center}
\end{figure}

 An important question is the behaviour of the relative deviation norm
 $\delta$ as a function of the {\em condition number} $\lambda/\epsilon$
 and of the order $n$.
 For fermionic simulations with the multi-bosonic algorithms the
 powers $\alpha=1,\half,{1 \over 4}$ are most interesting.
 I performed most of the tests with $\alpha={1 \over 4}$, which
 corresponds to the case of Majorana gluinos considered in
 refs.~\cite{TWO-STEP,HAM_MUN_COLL1,HAM_MUN_COLL2}.
 Several tests have also been performed with $\alpha=1$ and
 $\alpha=\half$, which are interesting, for instance, in numerical
 simulations of QCD with 2+1 light dynamical quarks.
 In present day fermion simulations condition numbers up to 
 $\lambda/\epsilon = {\cal O}(10^4)$ are in most cases sufficient.

 The behaviour of $\delta^2$ in the interesting range of parameters is
 illustrated by figures~\ref{fig01}, \ref{fig02} and \ref{fig03}.
 As is shown by the figures, the large-$n$ asymptotic behaviour of
 $\delta$ is well represented by the expected exponential decrease,
 already in the presented range or orders.
 In all three cases the observed behaviour is consistent with
\be \label{eq31}
\delta_n(\alpha;\epsilon/\lambda,1) \simeq
\exp\left\{ -C_\alpha n \sqrt{\epsilon/\lambda} \right\} \ ,
\ee
 where the constants $C_\alpha$ are approximately given by
 $C_{1/4} \simeq 2.3$, $C_{1/2} \simeq 2.2$ and $C_1 \simeq 2.0$,
 respectively.
 In fact, the fitted constants in figs.~\ref{fig01} and \ref{fig02} are
 close to these values for $\lambda/\epsilon=10^2$ and $10^3$, but
 still about 10-20\% higher for $\lambda/\epsilon=10^4$.
 In fig.~\ref{fig03} the  asymptotic behaviour is a reasonably good
 representation in the whole range for all three values of
 $\lambda/\epsilon$.
 
 Figures~\ref{fig01} to \ref{fig03} also demonstrate that for
 relatively low orders $\delta_n(\alpha;\epsilon/\lambda,1)$ evolve
 close to the upper limit $\delta_n(\alpha;0,1)$ given in
 eq.~(\ref{eq15}).
 This is required by the inequality in eq.~(\ref{eq12}), as long as
\be \label{eq32}
\frac{\alpha}{n+1+\alpha} \gg \sqrt{\epsilon/\lambda} \ .
\ee
 A rough estimate for the lowest $n$ where the relatively fast
 exponential decrease in (\ref{eq31}) sets in is given by the value of
 $n$ where the two sides of (\ref{eq32}) become equal. 


\section{Generalizations}\label{sec4}
 The great advantage of the quadratic polynomial approximation scheme
 defined in section~\ref{sec2} is its flexibility towards other
 functions and/or other regions of approximation.
 For instance, in the two-step multi-bosonic algorithm for fermions
 \cite{TWO-STEP} three different polynomial approximations are needed:
 the one considered up to now with a relatively low order $\bar{n}$,
 denoted by $\bar{P}(x)$, and two others with higher orders, realizing
 a better approximation.
 One of them has to approximate the function $x^{-\alpha}/\bar{P}(x)$
 therefore the relative deviation norm of $P(x)$ in eq.~(\ref{eq02})
 is replaced by
\be \label{eq33}
\delta \equiv \left\{ (\lambda - \epsilon)^{-1}\int_\epsilon^\lambda dx 
\left[ 1 - x^\alpha \bar{P}(x) P(x) \right]^2 \right\}^\half \ .
\ee
 The second higher order polynomial $\tilde{P}(x)$ is similarly defined.
 For a possible definition see ref.~\cite{TWO-STEP}, or alternatively, 
 take eq.~(\ref{eq33}) with $\alpha=0$ and
 $\bar{P} \to P,\; P \to \tilde{P}$ \cite{HAM_MUN_COLL1}.

 In order to find the minimum of $\delta$ in eq.~(\ref{eq33}) one can
 proceed similarly to sections~\ref{sec2} and \ref{sec3}.
 Denoting the coefficients of the polynomial $\bar{P}(x)$ by
 $\bar{c}_{\bar{\nu}}$, the coefficients of the quadratic form
 in (\ref{eq04}) are now replaced by
$$
V_\nu^{(\alpha)} = (\lambda - \epsilon)^{-1} 
\sum_{\bar{\nu}=0}^{\bar{n}} \bar{c}_{\bar{\nu}}\;
\frac{\lambda^{1+\alpha+\bar{n}-\bar{\nu}+n-\nu}
    -\epsilon^{1+\alpha+\bar{n}-\bar{\nu}+n-\nu}}
              {1+\alpha+\bar{n}-\bar{\nu}+n-\nu} \ ,
$$
\be \label{eq34} 
M_{\nu_1,\nu_2}^{(\alpha)} = (\lambda - \epsilon)^{-1}
\sum_{\bar{\nu}_1,\bar{\nu}_2=0}^{\bar{n}} 
\bar{c}_{\bar{\nu}_1}\bar{c}_{\bar{\nu}_2}\; \frac{
 \lambda^{1+2\alpha+2\bar{n}-\bar{\nu}_1-\bar{\nu}_2+2n-\nu_1-\nu_2} -
 \epsilon^{1+2\alpha+2\bar{n}-\bar{\nu}_1-\bar{\nu}_2+2n-\nu_1-\nu_2}}
          {1+2\alpha+2\bar{n}-\bar{\nu}_1-\bar{\nu}_2+2n-\nu_1-\nu_2}
\ .
\ee

 Another kind of generalization, besides (\ref{eq33}), is to consider
 different integration regions.
 For instance, the integration region can be split in several disjoint
 intervals.
 For integer power $\alpha$ the intervals can also be extended to
 negative $x$.
 As an interesting example, let us mention the case of integer $\alpha$
 with two intervals, one positive and another negative, lying
 symmetrically around zero.
 Such approximations can be useful for considering Hermitean fermion
 matrices with spectra symmetric around zero.

 Other optimized polynomial approximations are useful in different
 variants of multi-bosonic algorithms proposed by de Forcrand et al.
 \cite{FORCRAND}.
 In particular, in the non-Hermitean algorithms optimized approximations
 in the complex plane are required.
 In this case the interesting powers are twice as large as in the above
 Hermitean case, namely $\alpha=\half$ for Majorana fermions and,
 in general, $\alpha=N_f$ for $N_f$ equal-mass flavours.
 The relative deviation norm corresponding to eq.~(\ref{eq02}) is now
 defined by
\be \label{eq35}
\delta \equiv \left\{ \frac{
 \int_{\cal R}\, dx\,dy \left| 1 - (x+iy)^\alpha P(x+iy) \right|^2}
{\int_{\cal R}\, dx\,dy} \right\}^\half \ ,
\ee
 where ${\cal R}$ is a region in the complex plane containing the
 eigenvalue spectrum of the non-Hermitean fermion matrix.
 The coefficients of the polynomial $P(x+iy)$ are now, in general,
 complex.
 (Nevertheless, in special cases as in the explicit examples considered
 below, they can still be real.)
 The quadratic form for $\delta^2$ is also generally complex:
\be \label{eq36}
\delta^2 = 1 - \sum_{\nu=0}^n \left[ c_\nu^* V_\nu^{(\alpha)}
+ V_\nu^{(\alpha)*}c_\nu \right]
+ \sum_{\nu_1,\nu_2=0}^n c_{\nu_1}^* M_{\nu_1,\nu_2}^{(\alpha)} 
c_{\nu_2} \ ,
\ee
 where
$$
V_\nu^{(\alpha)} = \frac{
\int_{\cal R}\, dx\,dy (x-iy)^{\alpha+n-\nu}}{\int_{\cal R}\, dx\,dy}\ , 
$$
\be \label{eq37} 
M_{\nu_1,\nu_2}^{(\alpha)} = \frac{
 \int_{\cal R}\, dx\,dy (x-iy)^{\alpha+n-\nu_1}(x+iy)^{\alpha+n-\nu_2}}
{\int_{\cal R}\, dx\,dy}  \ .
\ee
 For the coefficients of the optimized polynomial eq.~(\ref{eq06}) still
 holds.

 In principle the shape of the complex region ${\cal R}$ can be quite
 arbitrary.
 However, for applications in multi-bosonic algorithms high polynomial
 orders are necessary, therefore it is advantageous to choose a region
 where the necessary integrals
\be \label{eq38}
{\cal I}_{k_1,k_2} \equiv
\int_{\cal R}\, dx\,dy (x-iy)^{k_1}(x+iy)^{k_2} \ ,
\ee
 with positive $k_1,k_2$, can be evaluated explicitly and the results
 are relatively simple.
 This is because the Maple procedures described in the previous section
 require high precision.

 In case of integer power $\alpha$ rectangular or elliptical shapes
 are well suited.
 Let us first consider a rectangle 
 $\{x \in [\epsilon,\lambda];\; y \in [-\delta,\delta]\}$ which is
 symmetric around the real axis.
 This is appropriate for fermion simulation algorithms because usually,
 if $\lambda$ is an eigenvalue of the fermion matrix then its complex 
 conjugate $\lambda^*$ also is.
 The integral in (\ref{eq38}), with positive integer $k_1,k_2$, is in
 this case the following:
$$
{\cal I}_{k_1,k_2} = 
\sum_{\kappa_1=0}^{k_1} \sum_{\kappa_2=0}^{k_2}
(-1)^{\half(\kappa_1-\kappa_2)}\;
\frac{\kappa_1!(k_1-\kappa_1)!\kappa_2!(k_2-\kappa_2)!}{k_1!k_2!} \cdot
$$
\be \label{eq39}
\frac{(\lambda^{1+k_1-\kappa_1+k_2-\kappa_2}
     -\epsilon^{1+k_1-\kappa_1+k_2-\kappa_2})}
              {(1+k_1-\kappa_1+k_2-\kappa_2)}\;
\frac{(\delta^{1+\kappa_1+\kappa_2}-(-\delta)^{1+\kappa_1+\kappa_2})}
             {(1+\kappa_1+\kappa_2)} \ .
\ee

 In case of an ellipse which is symmetric around the real axis with
 centre at $(x=\half(\lambda+\epsilon),y=0)$ and half-axes of lengths
 $\half(\lambda-\epsilon)$ and $\delta$ in the direction of the $x$- and
 $y$-axis, respectively, it is convenient to introduce the new
 integration variables
\be \label{eq40}
\xi = \frac{2x-\lambda-\epsilon}{\lambda-\epsilon} 
\ , \hspace{3em}
\eta = \frac{y}{\delta} \ . 
\ee
 The integral is then reduced to the unit circle ${\cal C}$ around the
 origin
\be \label{eq41}
{\cal I}_{k_1,k_2} =
\frac{(\lambda-\epsilon)\delta}{2} \int_{\cal C} d\xi\,d\eta 
\left( \half(\lambda+\epsilon) +
\half(\lambda-\epsilon)\xi -i\delta\eta \right)^{k_1}
\left( \half(\lambda+\epsilon) +
\half(\lambda-\epsilon)\xi +i\delta\eta \right)^{k_2}
\ee
 and a term-by-term evaluation gives:
$$
{\cal I}_{k_1,k_2} = {1 \over 4}(\lambda-\epsilon)\delta
\sum_{i_1,i_2,i_3=0}^{k_1} \delta_{i_1+i_2+i_3,k_1}
\sum_{j_1,j_2,j_3=0}^{k_2} \delta_{j_1+j_2+j_3,k_2}
$$
$$
\left(1+(-1)^{i_2+j_2}\right) \left(1+(-1)^{i_3+j_3}\right)
(-1)^{\half(i_3-j_3)}\; \frac{i_1!i_2!i_3!j_1!j_2!j_3!}{k_1!k_2!} \cdot
$$
\be \label{eq42}
\frac{\Gamma\left[\half(1+i_2+j_2)\right]
      \Gamma\left[\half(1+i_3+j_3)\right]}
     {\Gamma\left[1+\half(i_2+j_2+i_3+j_3)\right]}
\left(\frac{\lambda+\epsilon}{2}\right)^{i_1+j_1}
\left(\frac{\lambda-\epsilon}{2}\right)^{i_2+j_2} \delta^{i_3+j_3} \ .
\ee

 In the complex plane, up to now, we only considered $\alpha=integer$
 powers.
 For half-integer
\be \label{eq43}
\alpha \equiv A+\half
\ee
 the evaluation of the integrals on rectangles or ellipses becomes
 already cumbersome.
 A simple possibility is to go to the square-root complex plane 
 $(\xi,\eta)$ by the relations
\be \label{eq44}
x+iy = (\xi+i\eta)^2;\; \hspace{2em}
x = \xi^2-\eta^2,\; \hspace{2em} y=2\xi\eta \ .
\ee
 This transformation brings eq.~(\ref{eq35}) to the form
\be \label{eq45}
\delta \equiv \left\{ \frac{
 \int_{\cal R}\, d\xi\,d\eta (\xi^2+\eta^2) 
\left| 1 - (\xi+i\eta)^{1+2A} P[(\xi+i\eta)^2] \right|^2}
{\int_{\cal R}\, d\xi\,d\eta (\xi^2+\eta^2)} \right\}^\half \ .
\ee
 Here ${\cal R}$ is a region in the right half of the square-root
 complex plane such that its image under the mapping defined by
 (\ref{eq44}) covers the eigenvalue spectrum of the fermion matrix in
 the original $(x,y)$-plane.
 After this transformation one can take, for instance, rectangles or
 ellipses in the $(\xi,\eta)$-plane and apply formulae as (\ref{eq39})
 or (\ref{eq42}), respectively.

 Further generalizations are also possible:
 One can take, for instance, more complicated regions with boundaries
 given by polygons or ring shapes etc.
 Another possibility is to change the weight function $w(x)$ in
 the definition of the deviation norm (\ref{eq01})-(\ref{eq02}).
 Taking e.~g. $w(x) =1$ means to consider absolute deviations instead
 of relative ones.
 In multi-bosonic algorithms a special choice of the weight function can
 help to improve the quality of the approximations, for instance, by
 choosing $w(x)$ to be proportional to the average density of
 eigenvalues in the interval $[\epsilon,\lambda]$.


\section{Evaluating the polynomials: tests and comparisons}\label{sec5}
 The very high precision needed for obtaining the quadratically
 optimized polynomials does not mean that there is a similar difficulty
 in evaluating the polynomials.
 An extensive testing in the running project with dynamical gluinos
 in an SU(2) gauge theory showed \cite{HAM_MUN_COLL1,HAM_MUN_COLL2}
 that, in fact, there is no practical problem with the evaluation.
 Here only a small part of the performed tests are shown as
 represenative examples.
 The choice of parameters and lattice configurations is motivated by a
 work in progress \cite{HAM_MUN_COLL2}.
 For tests with the fermion matrix the (squared) even-odd
 preconditioned Hermitean fermion matrix is considered.

\begin{figure}
\begin{flushleft}
\epsfig{file=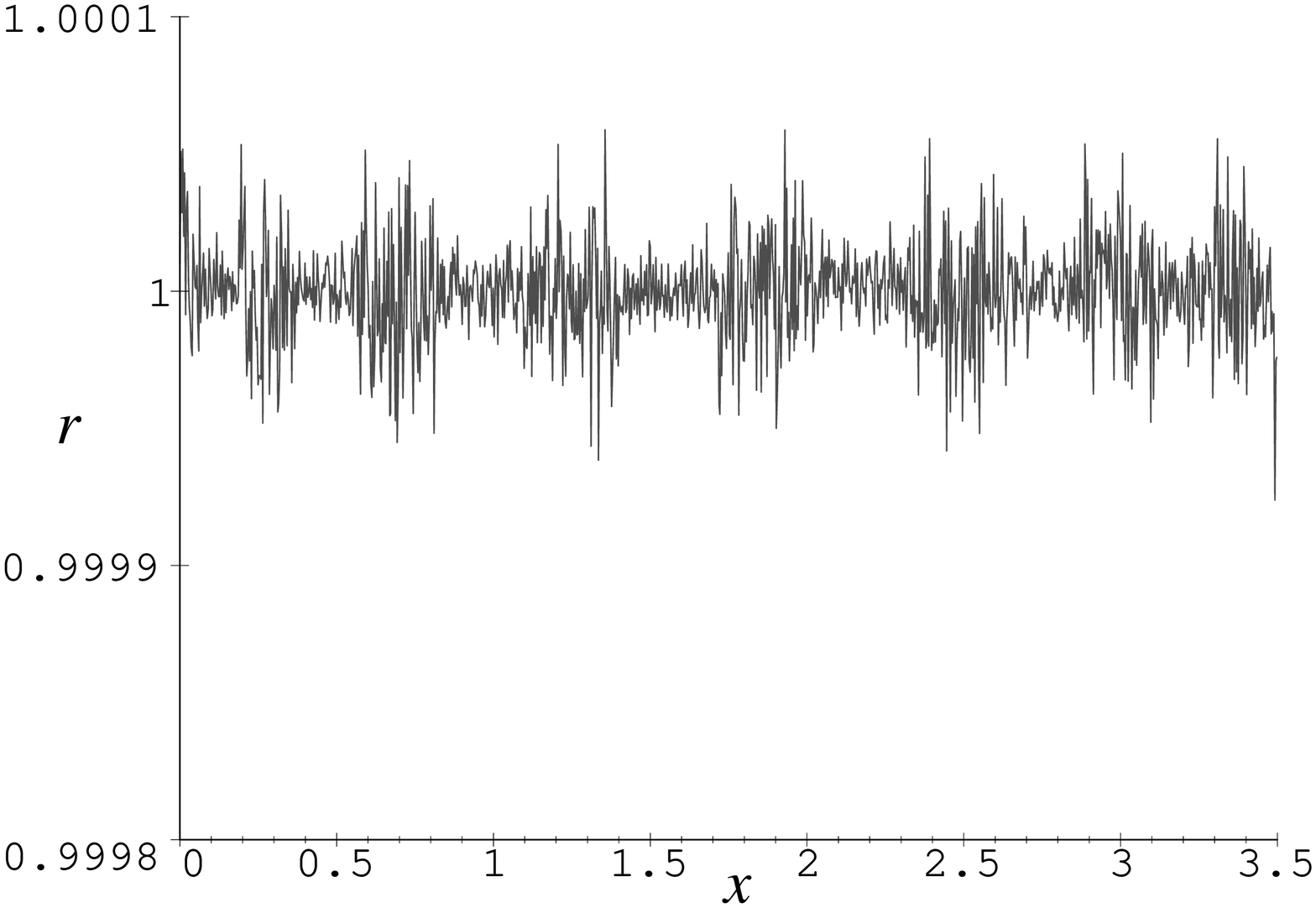,
        width=7.0cm,height=6.0cm,
        bbllx=20pt,bblly=50pt,bburx=600pt,bbury=620pt,
        angle=270}
\end{flushleft}\vspace{-8.0cm}
\begin{flushright}
\epsfig{file=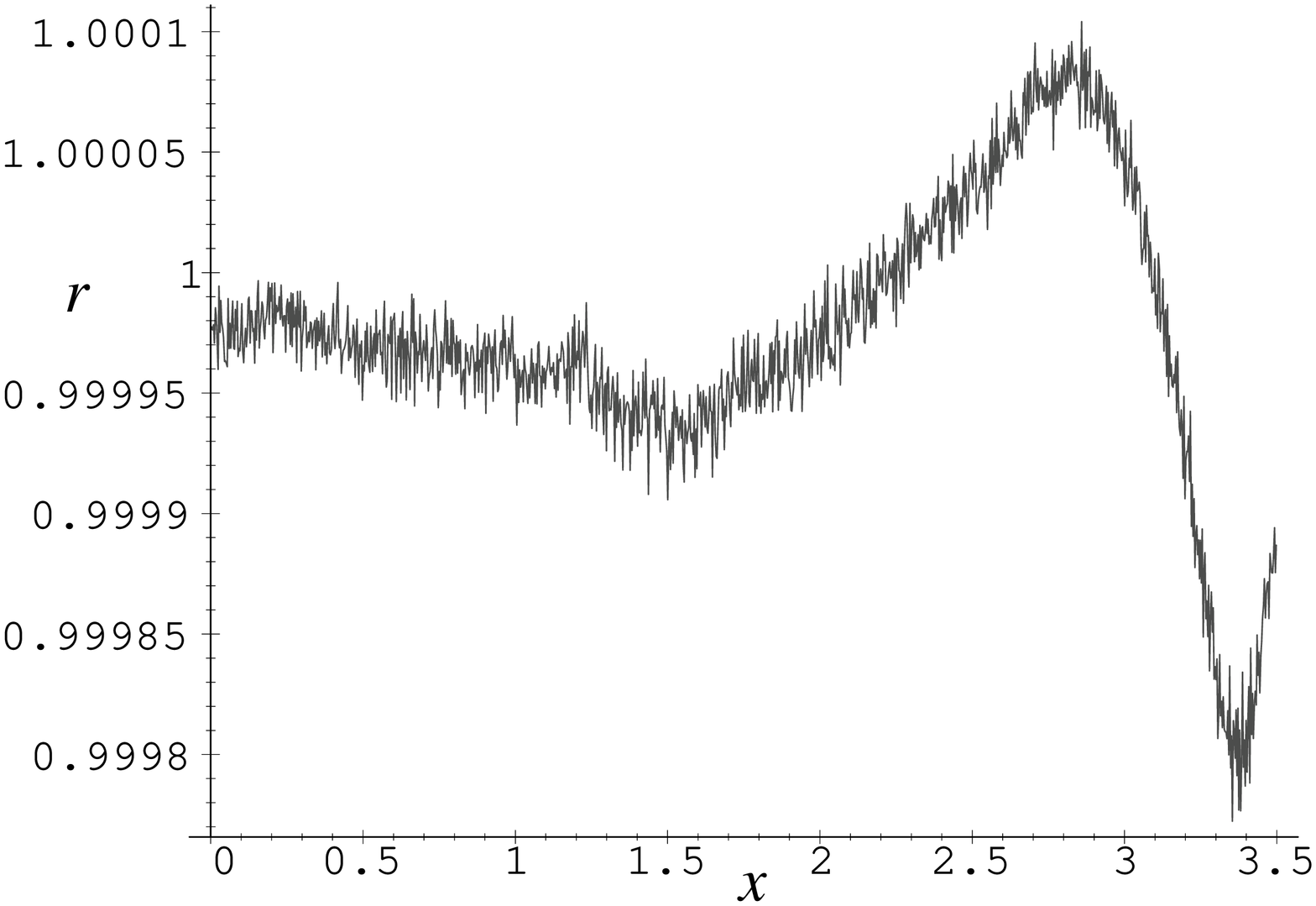,
        width=7.0cm,height=6.5cm,
        bbllx=20pt,bblly=150pt,bburx=600pt,bbury=720pt,
        angle=270}
\end{flushright}\vspace{-2.0em}
\begin{center}
\parbox{15cm}{\caption{ \label{fig04}
 The rounding errors with $Digits=6$ for the evaluation of a
 quadratically optimized polynomial, as specified in the text, with
 recurrence (left) and with root-factors in optimized order (right).
 (Note the small scale difference in the two halves.)}}
\end{center}
\end{figure}

 As a simple case, let us begin with an example for the evaluation of
 the polynomials on the real axis.
 We consider the polynomial approximation of order $n=72$ for
 $1/P_{48}(x)$ in the interval $[\epsilon,\lambda]=[0.0002,3.5]$,
 where $P_{48}(x)$ is a 48-th order optimized approximation of
 $x^{-1/4}/P_{16}(x)$ and $P_{16}(x)$ is a 16-th order optimized
 approximation of $x^{-1/4}$ in the same interval.
 The dependence of the result on the number of digits used is easily
 controlled by changing the $Digits$ parameter in Maple.
 In figure~\ref{fig04} the ratios of the results are shown for
 $Digits=6$ divided by $Digits=12$, which display the evaluation
 errors for $Digits=6$.
 In the left part of the figure the recursive evaluation discussed in
 section~\ref{sec2.1} is used, in the right part the root-product form
 in (\ref{eq09}) with optimized ordering of roots, as described in
 section~\ref{sec3}.
 As one can see, both evaluations give small errors of the order
 ${\cal O}(10^{-5})$-${\cal O}(10^{-4})$.
 The errors in case of the root-product form are somewhat larger and
 not completely uniform in the interval.

 For testing the evaluation of polynomials $P(X^2)$ of the square of the
 even-odd preconditioned Hermitean fermion matrix $X$ randomly chosen
 configurations were taken from updating series at $\beta=2.3$: at
 $K=0.185$ on $8^3 \cdot 16$ lattice and at $K=0.190$ on $6^3 \cdot 12$
 lattice.
 Approximations were considered for $x^{-1/4}$ with order $n_1$, for
 $x^{-1/4}/P_{n_1}$ with order $n_2$ and for $1/P_{n_2}$ with order
 $n_3$.
 In the two cases the orders were ($n_1=14,\,n_2=40,\,n_3=96$) and
 ($n_1=16,\,n_2=60,\,n_3=96$), respectively.
 The approximations were optimized in the intervals
 $[\epsilon,\lambda]=[0.001,3.4]$ and $[0.0002,3.5]$, respectively.
 The tests were carried out on the CrayT90 of HLRZ at J\"ulich with
 64-bit arithmetics.
 For estimating the rounding errors the polynomials were evaluated
 on random Gaussian starting vectors of unit length in three different
 ways and the components of the differences of result vectors were
 considered.
 
 The first way of evaluation was the root-product form in (\ref{eq09}).
 For defining the second evaluation, the pairs of complex conjugate
 roots were decomposed, as usual, according to
\be \label{eq46}
(X^2 - r)(X^2 - r^*) = (X - \rho_1  )(X - \rho_2  )
                       (X - \rho_1^*)(X - \rho_2^*)
\ee
 and the half-polynomial
\be \label{eq47}
P^{1/2}_n(X) \equiv \sqrt{c_{n0}} \prod_{j=1}^n (X-\rho_{nj})
\ee
 was taken, which gives
\be \label{eq48}
P_n(X^2) = P^{1/2}_n(X)^\dagger P^{1/2}_n(X) \ .
\ee
 In this second case, in the optimized ordering of roots the complex
 conjugate pairs were kept together, as mentioned in section~\ref{sec3}.
 The third way of evaluation was the recursive one discussed in 
 section~\ref{sec2.1}.
 The performed arithmetics are in these three cases completely
 different, hence the rounding errors are also different.
 The obtained averages for the real and imaginary parts of the
 components of difference vectors, which would be zero for infinite
 precision, are shown in table~\ref{tab01}.
 Since the rounding errors are independent for the pairs of evaluations
 considered, the individual roundind errors of an evaluation are smaller
 than those of the corresponding pairs.
 The numbers in the table show that
\be \label{eq49}
\sigma_1 > \sigma_2 > \sigma_3 \ .
\ee
 In general, all the three values are comfortably small.
 In fact, they are of the same order as the rounding errors in global
 sums of order ${\cal O}(1)$ quantities over the lattice.
 The standard deviations given in the table, which were obtained by
 repeating the calculation 100 times with different random starting
 vectors, show that the rounding errors depend little on the starting
 vector.
 Similarly, the particular gauge configuration chosen from the updating
 did not matter either.
 The dependence on the polynomial order and type is illustrated on the
 same configurations by table~\ref{tab02}.
\begin{table}[th]
\begin{center}
\parbox{15cm}{\caption{ \label{tab01}
 The average value of the real and imaginary parts of components of
 difference vectors $\sigma_{12},\,\sigma_{13},\,\sigma_{23}$ for three
 different evaluations of the second polynomials $P_{n_2}$, as defined
 in the text.
 The values in the table are given in units of $10^{-12}$.
 The digits in parentheses give standard deviations as observed for
 different starting vectors.}}
\end{center}
\vspace*{-1.7em}
\begin{center}
\begin{tabular}{| c | l | l | l |}\hline
lattice  &  \multicolumn{1}{c|}{$\sigma_{12}$}  
         &  \multicolumn{1}{c|}{$\sigma_{13}$}
         &  \multicolumn{1}{c|}{$\sigma_{23}$}          \\
\hline
$8^3 \cdot 16$  &  0.472(8)  & 0.434(9)  &  0.1569(7)    \\
\hline
$6^3 \cdot 12$  &  0.91(7)   &  0.88(8)  &   0.207(1)    \\
\hline\hline
\end{tabular}
\end{center}
\end{table}
\begin{table}[th]
\begin{center}
\parbox{15cm}{\caption{ \label{tab02}
 The same as table \protect{\ref{tab01}} for the three different
 polynomials considered.
 The case of root-product minus recursive evaluation is shown.}}
\end{center}
\vspace*{-1.4em}
\begin{center}
\begin{tabular}{| c | l | l | l |}\hline
lattice  &  \multicolumn{1}{c|}{$\sigma_{13}(P_{n_1})$}  
         &  \multicolumn{1}{c|}{$\sigma_{13}(P_{n_2})$}
         &  \multicolumn{1}{c|}{$\sigma_{13}(P_{n_3})$}  \\
\hline
$8^3 \cdot 16$  &  0.0402(3) & 0.434(9)  &  0.1243(4)    \\
\hline
$6^3 \cdot 12$  &  0.0536(7) & 0.88(8)   &  0.1473(9)    \\
\hline\hline
\end{tabular}
\end{center}
\end{table}

\begin{figure}
\begin{flushleft}
\epsfig{file=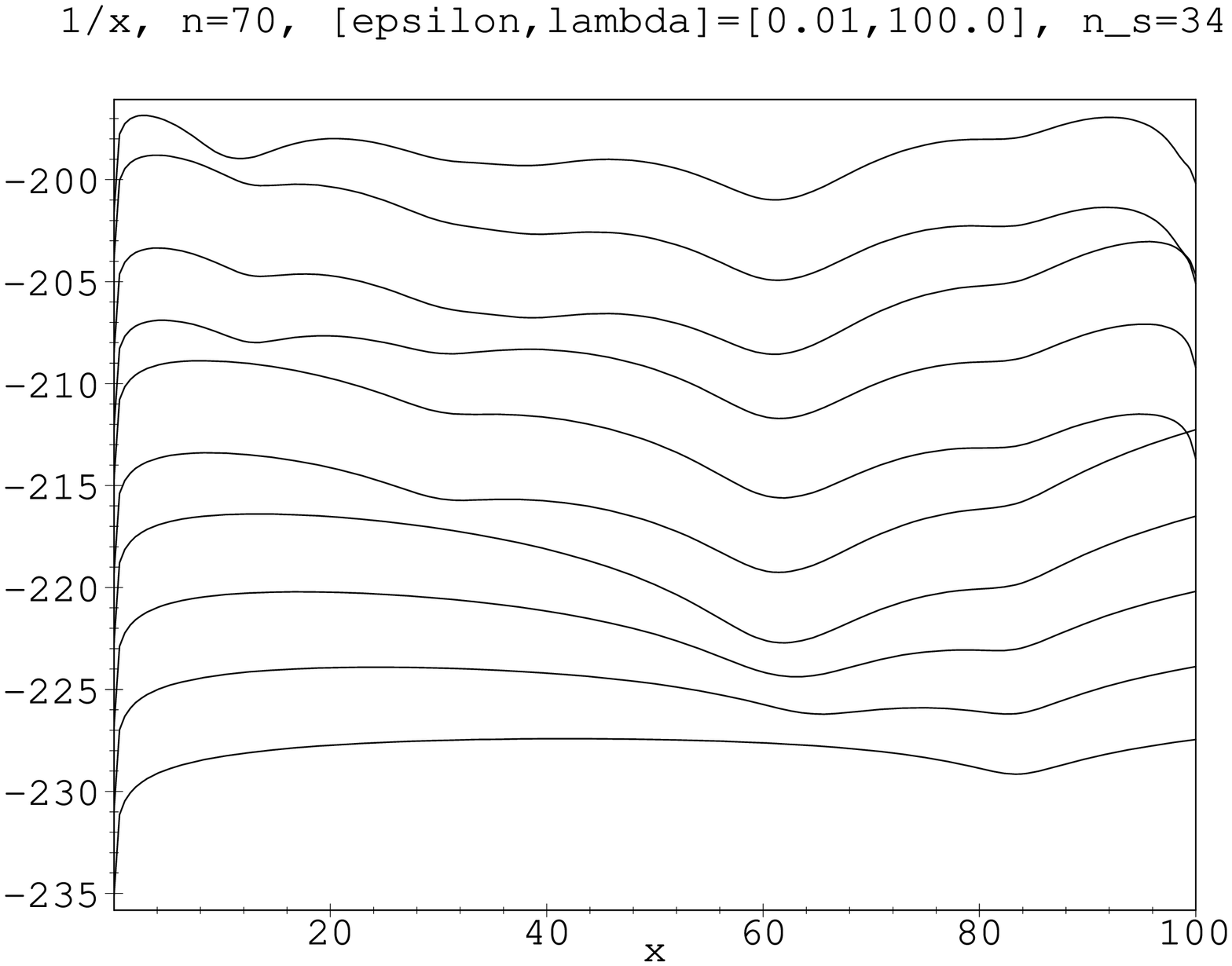,
        width=7.0cm,height=7.0cm,
        bbllx=20pt,bblly=50pt,bburx=600pt,bbury=620pt,
        angle=270}
\end{flushleft}\vspace{-8.0cm}
\begin{flushright}
\epsfig{file=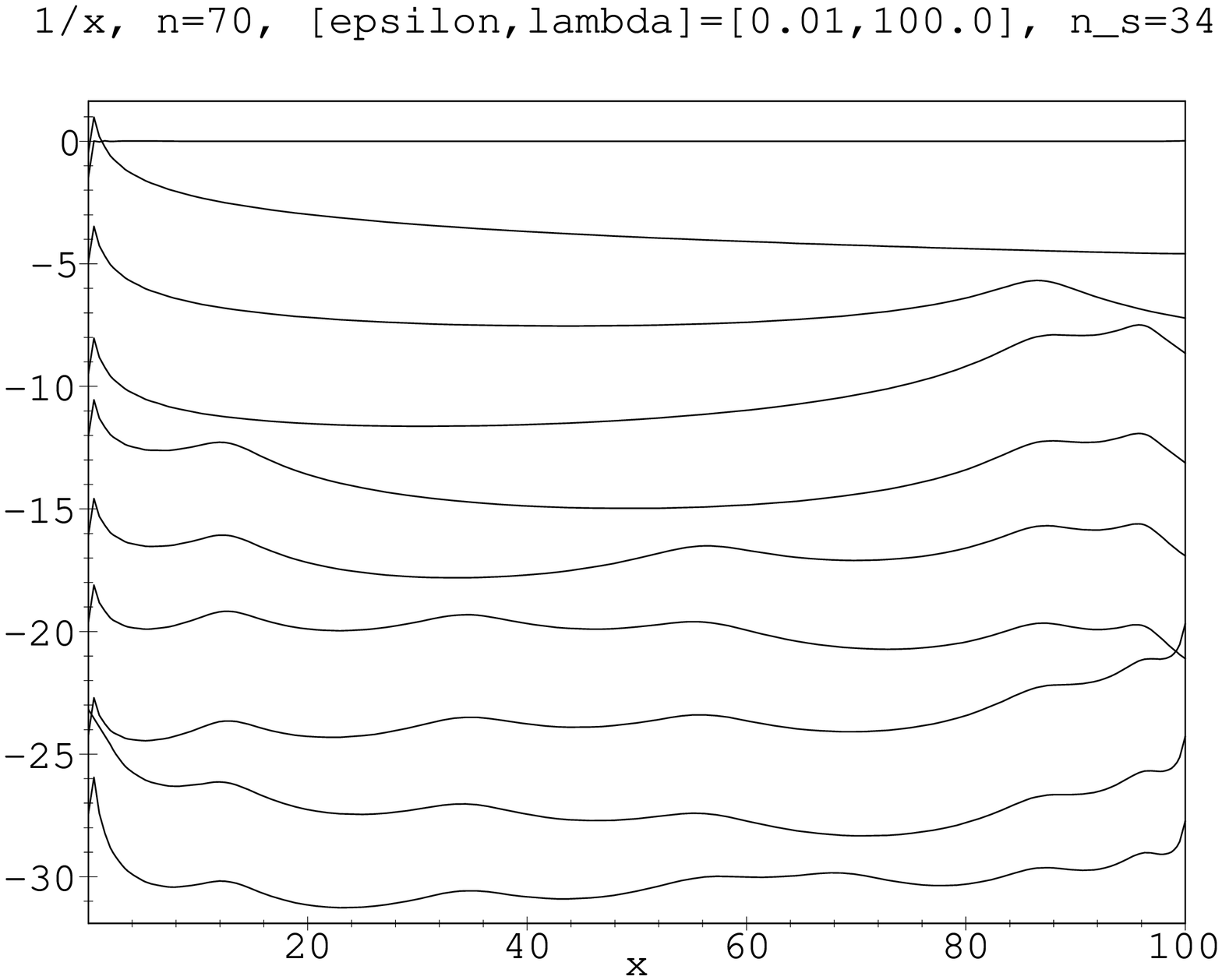,
        width=7.0cm,height=7.0cm,
        bbllx=20pt,bblly=150pt,bburx=600pt,bbury=720pt,
        angle=270}
\end{flushright}\vspace{-2.5em}
\begin{center}
\parbox{15cm}{\caption{ \label{fig05}
 The evolution of the partial products towards the function $1/x$ to be
 approximated in the interval $[\epsilon,\lambda]=[0.01,100.0]$.
 The order of the polynomial is $n=70$.
 The optimization of the root ordering is performed on $n_s=34$ points
 in the interval.
 The curves show the natural logarithms $\log|xP_p(x)|$ for the partial
 products $P_p(x)$ with $p=1,..,10$ (left) and $p=61,..,70$ (right).}}
\end{center}
\end{figure}

 Figure~\ref{fig04} and tables~\ref{tab01} and \ref{tab02} already show
 that in these typical cases the achieved optimization of root orderings
 is completely satisfactory.
 In any case, the best is to use the recurrence relations and not the
 root-factor products, if possible.
 The quality of root orderings can also be demonstrated by showing the
 $x$-dependence of the partial products of root-factors $P_p(x)$ for
 $p=1,2,\ldots,n$.
 This is shown in a case by figure~\ref{fig05}.
 Remember that during the optimization the depicted curves were tried to
 be kept as close to horizontal lines as possible.

\begin{figure}
\begin{center}
\epsfig{file=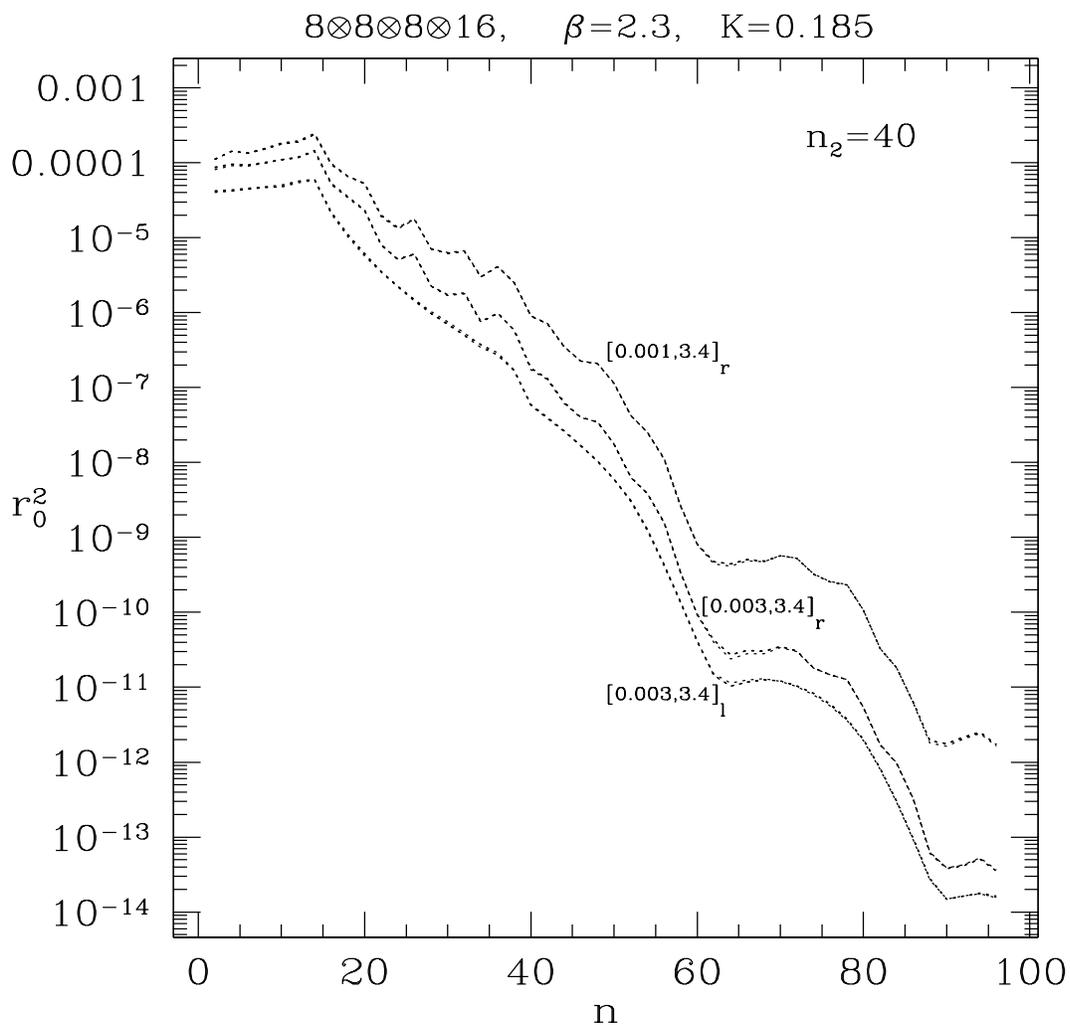,
        width=15.0cm,height=15.0cm,
        bbllx=40pt,bblly=100pt,bburx=620pt,bbury=720pt,
        angle=0}
\end{center}
\vspace{-4.0em}
\begin{center}
\parbox{15cm}{\caption{ \label{fig06}
 The length-square of the residue vector $r_0^2$ as a function of the
 polynomial order $n \equiv n_3^\prime$.
 The inverse of an optimized polynomial of order $n_2=40$ is calculated.
 The polynomials with order $n$ are optimized on the intervals
 $[\epsilon,\lambda]=[0.001,3.4]$ or $[0.003,3.4]$ and the starting
 vectors are random Gaussian ($r$) or local from a random starting point
 ($l$), as shown at the curves.}}
\end{center}
\end{figure}
 The third polynomial considered in the above examples (and in the
 two-step multi-bosonic algorithm) is the approximate inverse of the
 second one.
 In this case the independence of the polynomial coefficients on the
 final maximal order can be used to monitor the length of the residue
 vector of the inversion $r_0$:
\be \label{eq50}
r_0 \equiv | v_0 - P_{n_2}(X^2)P_{n_3^\prime}(X^2)v_0 |
\ee
 and stop the iteration if some required  precision is achieved.
 Here $v_0$ is the starting vector and $P_{n_3^\prime}$ is a polynomial
 with smaller order than $P_{n_3}$.
 In these tests, besides random Gaussian vectors of unit length also
 local starting vectors with a single non-zero component were
 considered from a randomly chosen site.  
 The interval of optimization was also changed.
 In figure~\ref{fig06}, which was obtained on an $8^3 \cdot 16$ 
 configration at $(\beta=2.3,K=0.185)$, the order of the second
 polynomial is $n_2 = 40$.
 As before, the curves depend only very little on the starting vector
 with a given type.
 Actually, the curves are superpositions for three different random
 starting vectors of the same type.

 This type of inversion for polynomials with values close to one
 compares rather favourably with more conventional ones as, for
 instance, conjugate gradient iteration.
 (Remember that the values of $P_{n_2}$ are close to one because it is
 an approximation to $x^{-1/4}/P_{n_1}(x)$.)
 In fact, in order to obtain the same precision in $r_0$, one needs 4
 CG iteration steps corresponding to $11n_2=440$ matrix multiplications
 with $X^2$, as compared to $n=90$ in the figure.
 (Here the calculation of the residues from (\ref{eq50}) is not
 counted, as it is not necessary for the inversion.)

\begin{figure}
\vspace{-1.0cm}
\begin{center}
\epsfig{file=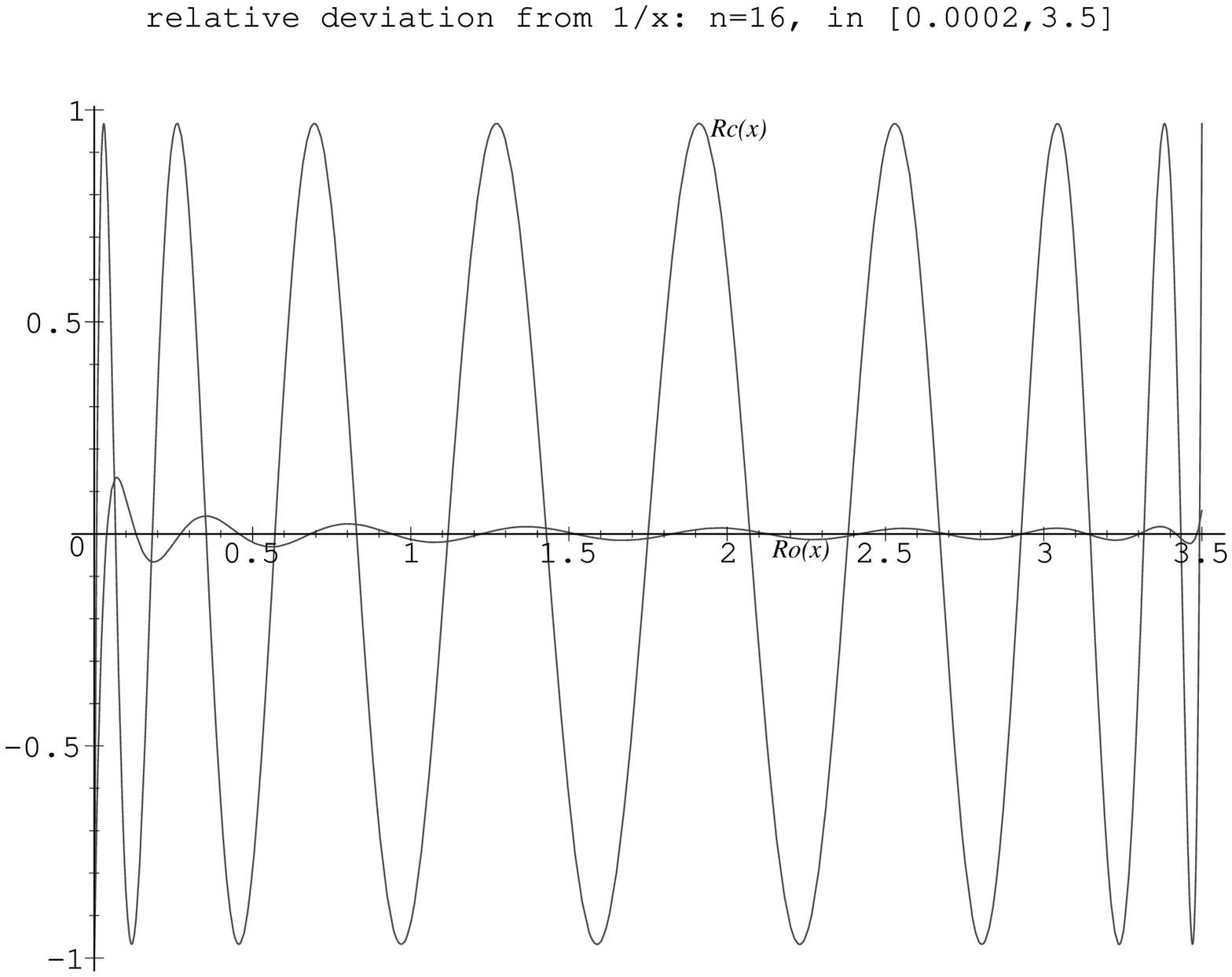,
        width=14.0cm,height=15.0cm,
        bbllx=20pt,bblly=50pt,bburx=600pt,bbury=670pt,
        angle=270}
\end{center}
\vspace{-3.0cm}
\begin{flushleft}
\epsfig{file=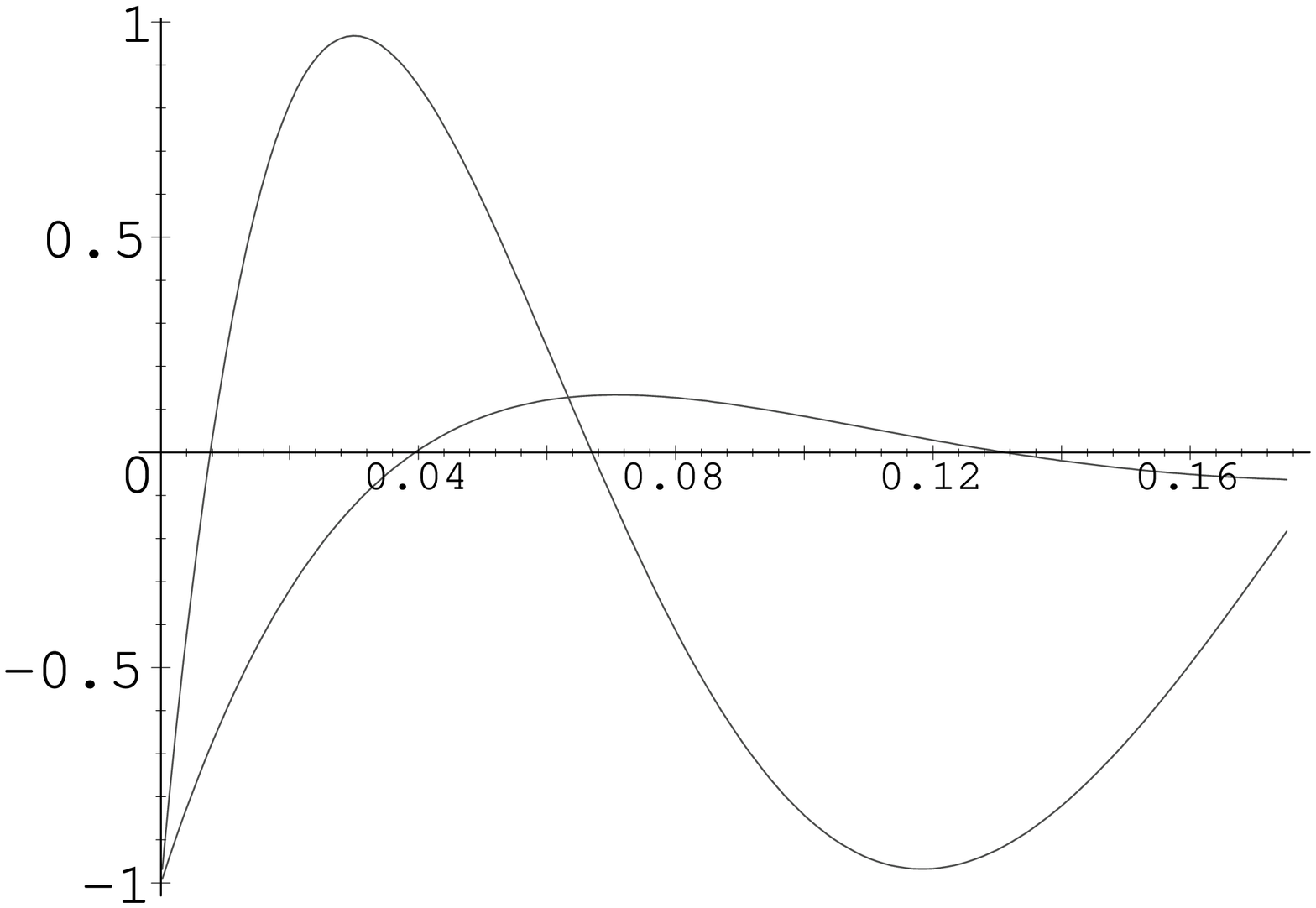,
        width=7.0cm,height=7.0cm,
        bbllx=20pt,bblly=10pt,bburx=600pt,bbury=580pt,
        angle=270}
\end{flushleft}\vspace{-8.0cm}
\begin{flushright}
\epsfig{file=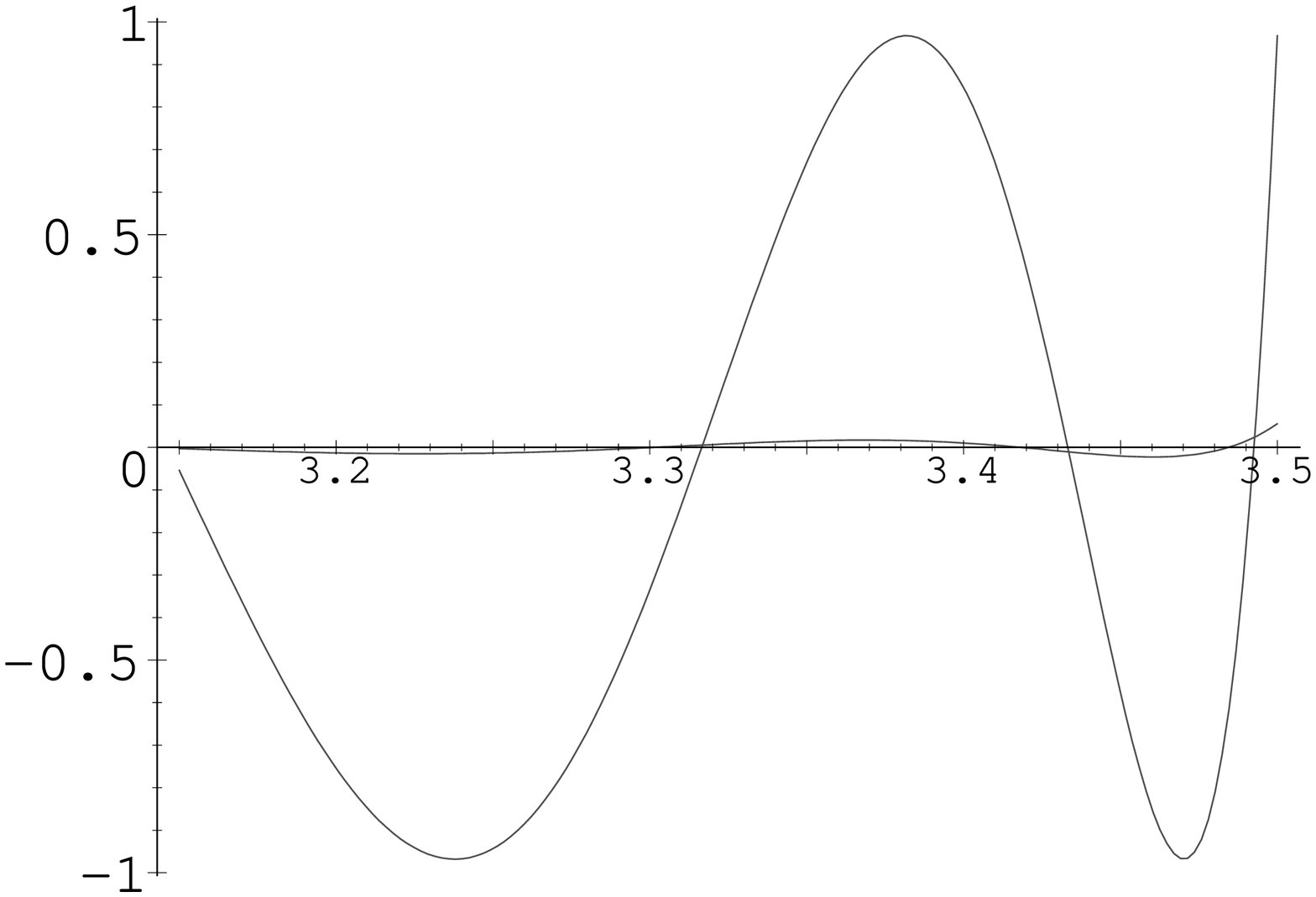,
        width=7.0cm,height=7.0cm,
        bbllx=20pt,bblly=110pt,bburx=600pt,bbury=680pt,
        angle=270}
\end{flushright}
\vspace{-3.0em}
\begin{center}
\parbox{15cm}{\caption{ \label{fig07}
 The comparison of the relative deviation $R(x) \equiv xP(x)-1$ for the
 Chebyshev polynomial $(Rc(x))$ and the quadratically optimized 
 polynomial $(Ro(x))$.
 In the example shown the approximation interval is $[0.0002,3.5]$ and
 the order $n=16$ is taken in both cases.
 The lower part of the figure zooms on the two ends of the interval.
 In the left lower corner the maximum deviation of the Chebyshev
 polynomial is smaller: $Rc(0.0002)=-0.968$ compared to
 $Ro(0.0002)=-0.991$.}}
\end{center}
\end{figure}

 In the special case of the function $x^{-1}$ the quadratically
 optimized polynomials may be compared to the Chebyshev polynomials
 used in \cite{LUSCHER}, which minimize the maximal relative deviation.
 The asymptotic behaviour of $\delta$ for the Chebyshev polynomials is
 also given by a formula as (\ref{eq31}) with $C_1 = 2$.
 However, the constants in front of the exponent are quite different.
 It turns out that at low orders the quadratically optimized polynomials
 are much better almost everywhere in the interval of approximation
 $[\epsilon,\lambda]$.
 There is only a very small piece near $\epsilon$ where the relative
 deviation of the Chebyshev polynomial of same order is a little bit
 better.
 This is illustrated by figure~\ref{fig07}.
 The advantage of quadratic optimization becomes larger for larger
 condition numbers $\lambda/\epsilon$.
 Concerning the roots: it turns out that in general the roots of the
 Chebyshev polynomials are much closer to the real axis that those of
 the quadratically optimized ones.
 For instance, in case of the polynomials in figure~\ref{fig07} the
 absolute values of the imaginary parts are an order of magnitude
 smaller.

 Of course, if really the maximal relative deviation matters then the
 Chebyshev polynomials are better by definition.
 In the two-step multi-bosonic algorithm
 \cite{TWO-STEP,HAM_MUN_COLL1,HAM_MUN_COLL2} there are two places where
 the quality of the polynomial approximations have an essential
 influence on the performance.
 First, one wants to achieve a good acceptance rate in the noisy
 correction step for a low order first polynomial and, second, one wants
 to perform the correction on random Gaussian vectors with second and
 third polynomials of order as low as possible.
 Concerning acceptance, the experience shows \cite{HAM_MUN_COLL2}
 that an acceptance about 90\% can be achieved on $8^3 \cdot 16$
 lattices as a whole at large condition numbers above $10^4$ with a
 quadratically optimized first polynomial of order $n_1 \leq 16$.
 In order to compare to Chebyshev polynomials, the lengths of the
 residue vectors
\be \label{eq51}
r_1 \equiv | v_0 - X^2 P_n(X^2)v_0 |
\ee
 were considered on random Gaussian and local starting vectors from a
 randomly chosen site.
 (In both cases the starting vectors had unit length.)
 In order to see the dependence on the starting vector, this was
 repeated 100 times for different $v_0$.
 The results on $8^3 \cdot 16$ and $6^3 \cdot 12$ configurations are
 shown in table~\ref{tab03} and \ref{tab04}, respectively.
 One can see that the corresponding values for quadratic optimization
 are by a factor 15-200 smaller.
 For instance, if one would require the same residues in case of
 Chebyshev approximation as for the lowest order quadratic optimizations
 shown in the tables, one would need roughly 80-th order instead of
 14-th order in table~\ref{tab03} and 150-th order instead of 16-th
 order in table~\ref{tab04}.
 These conclusions do not depend on the particular configuration.
 If one picks some other gauge configurations from the updating series,
 the numbers in tables~\ref{tab03} and \ref{tab04} do not change more
 than 20-30\% and the ratios between the left two and right two
 columns even less.
 The smaller polynomial orders result in non-trivial gains of
 performance in the two-step multi-bosonic algorithm, for instance,
 because the autocorrelations are roughly proportional to the order
 $n_1$ of the first polynomials.
\begin{table}[th]
\begin{center}
\parbox{15cm}{\caption{ \label{tab03}
 Comparison of the lengths of residue vectors $r_1$ defined in
 \protect{(\ref{eq51})} at different orders on a $8^3 \cdot 16$ lattice
 at ($\beta=2.3,K=0.185$).
 The starting vectors of unit length $v_0$ are either {\em random}
 Gaussian or {\em local} with a single non-zero component from a
 randomly chosen site.
 The digits in parentheses give standard deviations as observed for
 different starting vectors of the same type.}}
\end{center}
\begin{center}
\begin{tabular}{| c | l | l | l | l |}\hline
      & \multicolumn{2}{c|}{Chebyshev} 
      & \multicolumn{2}{c|}{quadratic}                         \\
\hline
order &  random  &  local  &  random  &  local                 \\
\hline
14  &  0.628(1)  &  0.441(2)  &  0.0417(6)  &  0.0242(12)      \\
\hline
40  &  0.562(1)  &  0.231(1)  & 0.0093(2)   &  0.00725(9)      \\ 
\hline
96  &  0.0569(2) &  0.0359(1) &  0.00132(2) &  0.00074(3)      \\ 
\hline\hline
\end{tabular}
\end{center}
\end{table}
\begin{table}[th]
\begin{center}
\parbox{15cm}{\caption{ \label{tab04}
 The same as table \protect{\ref{tab03}} on a $6^3 \cdot 12$ lattice at 
 ($\beta=2.3,K=0.190$).}}
\end{center}
\begin{center}
\begin{tabular}{| c | l | l | l | l |}\hline
      & \multicolumn{2}{c|}{Chebyshev} 
      & \multicolumn{2}{c|}{quadratic}                         \\
\hline
order &  random  &  local  &  random  &  local                 \\
\hline
16  &  1.145(4)  &  0.484(3)  &  0.039(2)   &  0.026(3)        \\
\hline
60  &  0.813(3)  &  0.343(2)  & 0.0072(11)  &  0.0051(9)       \\ 
\hline
96  &  0.483(2)  &  0.219(1)  &  0.0025(2)  &  0.0017(2)       \\ 
\hline
192 &  0.1017(3) &  0.0538(4) &  0.00081(8) &  0.00050(7)      \\ 
\hline\hline
\end{tabular}
\end{center}
\end{table}

\newpage
\section{Summary}\label{sec6}
 In this paper the quadratically optimized polynomial approximations
 necessary for multi-bosonic algorithms of fermion simulations
 are considered.

 In section~\ref{sec2} the definitions and basic properties of this
 scheme are introduced in a simple case, namely, optimization of the
 relative quadratic deviation from the function
 $x^{-\alpha}\; (\alpha > 0)$ in a positive interval
 $x \in [\epsilon,\lambda]$ $(0 \leq \epsilon < \lambda)$.
 The expansion in suitably defined orthogonal polynomials is also
 considered.
 This allows for a recursive evaluation of the polynomials, without the
 knowledge of the roots.

 An algorithm in the algebraic manipulation language Maple V is
 described in section~\ref{sec3}. 
 With its help the coefficients and roots of the optimized polynomials
 can be determined, together with an optimal ordering of the roots
 for the application of the polynomial of the fermion matrix in product
 form with floating point arithmetics.

 In section~\ref{sec4} generalizations are discussed which are
 necessary in different variants of multi-bosonic algorithms: 
 approximations with products of polynomials and extension of the
 region of approximation from the real axis to the complex plane.

 In general, the calculations with the given algorithms can easily be
 performed on modern workstations for polynomial orders as high as
 $n=100-300$.
 These are typically the maximal orders one needs in present day
 numerical simulations of fermionic quantum field theories.
 This is illustrated by figures~\ref{fig01}, \ref{fig02} and
 \ref{fig03} where very good approximations with small $\delta$ are
 achieved already below $n=100$.
 Experience in SU(2) Yang-Mills theory with gluinos tells
 \cite{HAM_MUN_COLL1,HAM_MUN_COLL2} that for the interesting
 values of $\lambda/\epsilon \simeq {\cal O}(10^3)$, with rather high
 statistics, practically no deviation of the expectation values can be
 observed already for $\delta^2 \simeq {\cal O}(10^{-6})$.
 In these figures the simple polynomials defined in section~\ref{sec2}
 are considered, but very similar results hold also for the other two
 polynomials needed in the two-step multi-bosonic algorithm 
 \cite{TWO-STEP,HAM_MUN_COLL1}, which are discussed in
 section~\ref{sec4}.

 There is no principal obstacle to extend the calculations to higher
 orders, but then the requirements on computer power increase and
 further improvements of the Maple algorithms are welcome.

 Detailed tests for the evaluation of the optimized polynomials are
 shown in section~\ref{sec5} in typical situations relevant in
 numerical simulations with dynamical fermions.
 These are based on experience gained in the running project with
 dynamical gluinos in an SU(2) gauge theory
 \cite{HAM_MUN_COLL1,HAM_MUN_COLL2}.
 A comparison with Chebyshev polynomials in the special case of the
 function $x^{-1}$ is favourable for quadratic optimization.

 The main advantage of the quadratic (or least-squares) optimization is
 its generality and flexibility concerning the choice of functions and
 regions for the approximation.
 This is welcome in present and future large scale numerical simulations
 with dynamical fermions.

\newpage                                                    
{\large\bf Acknowledgements} 

\vspace{5mm}\noindent
 It is a pleasure to thank Martin L\"uscher for helpful comments.
 My collaborators in the dynamical gluino project, especially Klaus
 Spanderen and J\"org Westphalen, helped a lot with their comments. 
 The test gauge configurations were kindly provided from running
 updating series by Klaus Spanderen.


\vspace*{3em}
\begin{thebibliography}{99}
%
\bibitem{MONMUN}
I. Montvay, G. M\"unster,
{\em Quantum Fields on a Lattice}, Cambridge University Press, 1994.
%
\bibitem{LUSCHER}
M. L\"uscher,
Nucl. Phys. B418 (1994) 637.
%
\bibitem{TWO-STEP}
I. Montvay,
Nucl. Phys. B466 (1996) 259.
%
\bibitem{NOISY}
A.D. Kennedy, J. Kuti,
Phys. Rev. Lett. \underline{54} (1985) 2473;  \\
A.D. Kennedy, J. Kuti, S. Meyer, B.J. Pendleton,
Phys. Rev. \underline{D38} (1988) 627.
%
\bibitem{CHEBYSHEV}
L. Fox, I.B. Parker,
{\em Chebyshev Polynomials in Numerical Analysis,} Oxford University
Press, 1968.
%
\bibitem{BUNK}
B. Bunk,
talk given at the conference Lattice '97 in Edinburgh.
%
\bibitem{LAGUERRE}
J.H. Wilkinson,
{\em  Rounding Errors in Algebraic Processes,} 
London,1963.
%
\bibitem{HAM_MUN_COLL1}
G. Koutsoumbas, I. Montvay, A. Pap, K. Spanderen, D. Talkenberger,
J. Westphalen,
hep-lat/9709091.
%
\bibitem{HAM_MUN_COLL2}
I. Montvay, K. Spanderen, J. Westphalen,
in preparation.
%
\bibitem{FORCRAND}
A. Borici, Ph. de Forcrand,
Nucl. Phys. \underline{B454} (1995) 645; 
Nucl. Phys. B (Proc. Suppl.) \underline{47} (1996) 800; \\
A. Borrelli, Ph. de Forcrand, A. Galli,
Nucl. Phys. \underline{B477} (1996) 809; \\
A. Galli, Ph. de Forcrand,
Nucl. Phys. B (Proc. Suppl.) \underline{53} (1997) 956; \\
C. Alexandrou, A. Borici, A. Feo, Ph. de Forcrand, A. Galli, 
F. Jegerlehner, T. Takaishi,
Nucl. Phys. B (Proc. Suppl.) \underline{53} (1997) 435.
%
\end{thebibliography}
\end{document}